\documentclass[aps,prd,twocolumn,superscriptaddress,showpacs,preprintnumbers,amsmath,amssymb]{revtex4-2}

\usepackage[dvips]{graphicx} 
\usepackage{dcolumn}  
\usepackage{xcolor}
\usepackage{hyperref}
\hypersetup{breaklinks=true,colorlinks=true,allcolors=blue}
\usepackage{orcidlink}
\usepackage{multirow}

\renewcommand{\arraystretch}{1.1}

\begin{document}


\title{ \quad\\[1.0cm] Study of the lineshape of $X(3872)$ using $B$ decays to $D^0\overline{D}{}^{*0}K$}

\noaffiliation
  \author{H.~Hirata\,\orcidlink{0000-0001-9005-4616}} 
  \author{T.~Iijima\,\orcidlink{0000-0002-4271-711X}} 
  \author{Y.~Kato\,\orcidlink{0000-0001-6314-4288}} 
  \author{K.~Tanida\,\orcidlink{0000-0002-8255-3746}} 
  \author{I.~Adachi\,\orcidlink{0000-0003-2287-0173}} 
  \author{J.~K.~Ahn\,\orcidlink{0000-0002-5795-2243}} 
  \author{H.~Aihara\,\orcidlink{0000-0002-1907-5964}} 
  \author{S.~Al~Said\,\orcidlink{0000-0002-4895-3869}} 
  \author{D.~M.~Asner\,\orcidlink{0000-0002-1586-5790}} 
  \author{H.~Atmacan\,\orcidlink{0000-0003-2435-501X}} 
% ...
\collaboration{The Belle Collaboration}


\begin{abstract}
We present a study of the $X(3872)$ lineshape in the decay $B \to X(3872)K\to D^0\overline{D}{}^{*0}K$ using a data sample of $772\times 10^6$ $B\overline{B}$ pairs collected at the $\Upsilon(4S)$ resonance with the Belle detector at the KEKB asymmetric-energy $e^+e^-$ collider.
The peak near the threshold in the $D^0\overline{D}{}^{*0}$ invariant mass spectrum is fitted using a relativistic Breit-Wigner lineshape.
We determine the mass and width parameters to be $m_{\rm BW} = 3873.71 ^{+0.56}_{-0.50} ({\rm stat}) \pm0.13 ({\rm syst}) ~{\rm MeV}/c^2$ and $\Gamma_{\rm BW} = 5.2 ^{+2.2}_{-1.5} ({\rm stat}) \pm 0.4 ({\rm syst})~{\rm MeV}$, respectively.
The branching fraction is found to be ${\cal B} (B^+\to X(3872)K^+) \times {\cal{B}}(X(3872) \to D^0\overline{D}{}^{*0}) = (0.97 ^{+0.21}_{-0.18} ({\rm stat}) \pm 0.10 ({\rm syst})) \times 10^{-4}$.
The signal from $B^0$ decays is observed for the first time with $5.2\sigma$ significance, and the ratio of branching fractions between charged and neutral $B$ decays is measured to be ${\cal B}(B^0\to X(3872)K^0)/{\cal B}(B^+ \to X(3872)K^+) = 1.34^{+0.47}_{-0.40} ({\rm stat}) ^{+0.10}_{-0.12} ({\rm syst})$.
The peak is also studied using a Flatt\'{e} lineshape.
We determine the lower limit on the $D\overline{D}{}^{*}$ coupling constant $g$ to be $0.075$ at 95\% credibility in the parameter region where the ratio of $g$ to the mass difference from the $D^0\overline{D}{}^{*0}$ threshold is equal to $-15.11~{\rm GeV}^{-1}$, as measured by LHCb.
\end{abstract}

\maketitle

\tighten

{\renewcommand{\thefootnote}{\fnsymbol{footnote}}}
\newcommand{\Flatte}{Flatt\'{e} }
\newcommand{\BABAR}{{\sl BABAR} }
\newcommand{\red}[1]{\textcolor{red}{#1}}
\newcommand{\blue}[1]{\textcolor{blue}{#1}}
\newcommand{\violet}[1]{\textcolor{violet}{#1}}
\newcommand{\teal}[1]{\textcolor{teal}{#1}}

\setcounter{footnote}{0}

\section{introduction}
The charmonium-like $X(3872)$ state, also known as $\chi_{c1}(3872)$~\cite{PDG2020}, was discovered by the Belle experiment as a narrow peak in the vicinity of the $D^0\overline{D}{}^{*0}$ threshold in the $J/\psi \pi^+ \pi^-$ invariant mass distribution in exclusive $B^+ \to J/\psi \pi^+ \pi^- K^+$ decays~\cite{XDiscoveryBelle}.
Its existence has been confirmed by multiple experiments: D0~\cite{XDiscoveryD0}, {\sl BABAR}~\cite{XDiscoveryBABAR}, CDF~\cite{XDiscoveryCDF}, LHCb~\cite{XDiscoveryLHCb}, and BESIII~\cite{XDiscoveryBESIII}.
In addition to the $J/\psi \pi^+ \pi^-$ decay, other decays such as $J/\psi \omega$~\cite{JpsiomegaDiscovery}, $J/\psi \gamma$, $\psi(2S) \gamma$~\cite{radiativeDecaysDiscovery}, $D^0\overline{D}{}^{*0}$~\cite{BaBarDDstar, Aushev2010}, $D^0\overline{D}{}^{0}\pi^0$~\cite{Gokhroo2006}, and $\pi^0\chi_{c0}$~\cite{chicJpi0Discovery} have been observed.
The $X(3872)$ quantum numbers $J^{PC}$ have been determined to be $1^{++}$~\cite{LHCbXJpc1, LHCbXJpc2}.
Various interpretations such as a loosely bound state~\cite{molecular1, molecular2, molecular3, molecular4}, an admixture of a molecular state and a pure charmonium resonance~\cite{admixtureCCandMolecule}, a tetraquark~\cite{tetraquark}, and a cusp at the $D^0\overline{D}{}^{*0}$ threshold~\cite{Hanhart2007, Braaten2007, Bugg2008} have been proposed, and the structure of the state remains uncertain. 
Measurement of the lineshape in various decay modes can help to discriminate among different choices for the structure.
In this paper, we examine two models for the lineshape in the decay to $D^0\overline{D}{}^{*0}$: a Breit-Wigner, and a Flatt\'{e}-inspired parametrization.\par

The $X(3872)$ peak has already been analyzed with the Breit-Wigner lineshape commonly used for resonance states.
Based on the analyses of the decays including $J/\psi$, the mass is $3871.65\pm0.06~{\rm MeV}/c^2$ and the width is $1.19\pm0.21~{\rm MeV}$~\cite{PDG2020}, with two measurements at the LHCb experiment~\cite{LHCbBRJpsipipi2020, LHCbFlatte} contributing significantly to these averages.
Analyses of the decay to $D^0\overline{D}{}^{*0}$ based on the Breit-Wigner lineshape tend to yield a higher mass and a larger width,
with the width measurement subject to large uncertainties~\cite{BaBarDDstar, Aushev2010}.
Discrepancies in the lineshape between the decays to the $J/\psi \pi^+\pi^-$ and $D^0\overline{D}{}^{*0}$ final states can arise near the threshold due to coupled-channel effects~\cite{Hanhart2007}.
This may be significant for the $X(3872)$, as the observed mass coincides with the $D^0\overline{D}{}^{*0}$ threshold of $3871.69 \pm 0.10~{\rm MeV}/c^2$, and a $1^{++}$ state can couple to the $D^0\overline{D}{}^{*0}$ channel in S-wave.
One model to account for coupled-channel effects is the Flatt\'{e}-inspired parametrization~\cite{Hanhart2007, Kalashnikova2009}, a Breit-Wigner model with an explicit expression for the energy-dependent partial width.
At LHCb, an analysis of the $J/\psi \pi^+ \pi^-$ invariant mass distribution was performed using this Flatt\'e-inspired model~\cite{LHCbFlatte}.
It is difficult to determine all of the parameters using only this distribution, due to a scaling behavior in which the lineshape near the threshold does not change under a linear transformation of four of the five parameters~\cite{LHCbFlatte, Baru2005}.
To determine all the parameters, it is important to analyze not only the $J/\psi \pi^+ \pi^-$ decay but also the $D^0\overline{D}{}^{*0}$ decay, as proposed in the theoretical analysis~\cite{Kalashnikova2009}.
By analyzing the $D^0\overline{D}{}^{*0}$ decay, we aim to provide more information on the lineshape, and in particular on the coupling strength of $X(3872) \to D^0\bar{D}^{*0}$.

In this paper, we present a study of the $X(3872)$ lineshape using a sample of $X(3872)\to D^0\overline{D}{}^{*0}$ candidates produced in the exclusive decay $B \to D^0\overline{D}{}^{*0}K$ using the full Belle dataset.
There have been three previous studies~\cite{Gokhroo2006, BaBarDDstar, Aushev2010}.
Reference~\cite{Gokhroo2006} is an analysis of the $B \to D^0\overline{D}{}^0\pi^0 K$ decay at Belle, and
Refs~\cite{BaBarDDstar, Aushev2010} are analyses of the $B \to D^0\overline{D}{}^{*0}K$ decays at \BABAR and Belle, respectively.
The latter two analyses apply a $D^{*0}$ selection and a mass-constrained fit to the $D^{*0}$ candidates.
While this has the advantage of improving the signal-to-noise ratio, it has the disadvantage of disallowing entries below the $D^0\overline{D}{}^{*0}$ threshold, which is important for studying the structure.
Given the limited size of our data sample, we adopt a similar technique to the latter analyses, \emph{i.e.}\ subtracting the reconstructed $D^{*0}$ mass and adding the nominal mass.
The disadvantage of requiring the $D^{*0}$ is partially compensated for by analyzing the \Flatte model, in which we can obtain a lineshape reflecting poles of the scattering amplitude.
Compared to Refs~\cite{BaBarDDstar, Aushev2010}, additional $D^0$ decay modes are included, increasing the efficiency to reconstruct $D^0\overline{D}{}^{*0}$ decays.
Throughout this paper, charge conjugation is always included.
We do not distinguish $D^0\overline{D}{}^{*0}$ from $\overline{D}{}^0D^{*0}$ unless otherwise indicated.\par

The paper is organized as follows.
In Sec.~\ref{secID:detector}, the Belle detector and data set are described.
In Sections~\ref{secID:selection} and~\ref{secID:fitmodel}, the event selection and the fitted model are presented.
In Sec.~\ref{secID:fitresult}, the results of fitting the data with the relativistic Breit-Wigner model and the \Flatte model are presented.
Section~\ref{secID:conclusion} contains a discussion of the results, and the conclusions of the paper.\par

\section{Detector and Data Set}
\label{secID:detector}
We use a data sample of $772\times 10^6$ $B\overline{B}$ pairs,
collected at a center-of-mass energy of $\sqrt{s}=10.58~{\rm GeV}$, corresponding to the $\Upsilon(4S)$ resonance, 
with the Belle detector at the KEKB asymmetric-energy $e^+e^-$ collider~\cite{KEKB, achivementKEKB}.
The Belle detector is a large-solid-angle magnetic
spectrometer that consists of a silicon vertex detector (SVD),
a 50-layer central drift chamber (CDC), an array of
aerogel threshold Cherenkov counters (ACC),  
a barrel-like arrangement of time-of-flight
scintillation counters (TOF), and an electromagnetic calorimeter
comprised of CsI(Tl) crystals (ECL) located inside 
a super-conducting solenoid coil that provides a 1.5~T
magnetic field.  An iron flux-return located outside of
the coil is installed to detect $K_L^0$ mesons and to identify
muons (KLM).  The detector
is described in detail elsewhere~\cite{Belle, physicsAchivementBelle}.\par
To determine the event selection and the detector response, we use a sample of Monte Carlo (MC) simulated events generated using the \texttt{EvtGen} event generator~\cite{evtgen}.
The detector response is simulated using the \texttt{GEANT3} package~\cite{geant3}.

\section{Event Selection}
\label{secID:selection}

The event selection is determined using the MC samples in two steps.
First, the selection criteria for the final-state particles are determined based on our previous studies~\cite{Gokhroo2006, Aushev2010}.
Second, the selection criteria for the intermediate-state particles are optimized by maximizing the figure-of-merit $S/\sqrt{S+B}$, where $S$ and $B$ are the estimated numbers of signal and background events, respectively.
The resulting selection is described below.
\par

Tracks are selected using vertex information measured by the tracking system.
A track candidate is accepted if its distance along the detector axis 
from the point of closest approach to the interaction point 
is less than 4.0~cm,
and its distance transverse to the detector axis
is less than 1.0~cm.
These requirements are not imposed for tracks in $K_S^0 \to \pi^+ \pi^-$ candidates.
In addition, pion and kaon candidates are selected using likelihoods ${\cal L}_{\pi}$ and ${\cal L}_K$ based on the time-of-flight measured by the TOF, the number of Cherenkov photons detected by the ACC, and the ionization loss in the CDC.
Tracks with a likelihood ratio ${\cal L}_\pi/({\cal L}_\pi + {\cal L}_K) > 0.1$ are used as charged pion candidates, and tracks with ${\cal L}_\pi/({\cal L}_\pi + {\cal L}_K) < 0.9$ are used as charged kaon candidates.
The hadron identification efficiency is approximately 97\% for both pions and kaons.
Tracks satisfying ${\cal L}_e/({\cal L}_e+{\cal L}_{\tilde{e}}) > 0.95$ are identified as electrons and eliminated.
Here, ${\cal L}_e$ and ${\cal L}_{\tilde{e}}$ are distinct likelihoods for the electron and non-electron hypotheses,
based on ECL, tracking, and other information. 
The particle identification is described in detail elsewhere~\cite{BellePID}.
\par

$K_S^0$ candidates are reconstructed from charged pion pairs with opposite charges.
The $\pi^+ \pi^-$ invariant mass is required to agree with the known $K_S^0$ mass~\cite{PDG2020} within $7~{\rm MeV}/c^2$ ($\approx 3.6\sigma$ of the resolution).
Candidates are selected using a neural network classifier~\cite{neurobayes} with various kinematic variables as input.
To improve the four-momentum resolution, a mass- and vertex-constrained fit is applied.\par

Photon candidates are reconstructed from ECL clusters with no matching charged tracks.
Candidates are selected based on the ratio, $E_9/E_{25}$, of the energy deposited in the $3\times3$ array of crystals centered on the crystal with the highest energy deposition to that in the $5\times 5$ array: we require $E_9/E_{25} > 0.8$.\par

Neutral pions are reconstructed from photon pairs.
The photons are required to have energy greater than $30~{\rm MeV}$ in the barrel region or $50~{\rm MeV}$ in the endcaps.
The $\gamma \gamma$ invariant mass is required to agree with the $\pi^0$ nominal mass~\cite{PDG2020} within $12~{\rm MeV}/c^2$.
This mass window corresponds to 92\% signal efficiency.
A mass-constrained fit is applied to improve the momentum resolution.\par

$D^0$ candidates are reconstructed in six decay modes: $K^- \pi^+$, $K^- \pi^+ \pi^0$, $K^-\pi^+\pi^-\pi^+$, $K_S^0 \pi^+ \pi^-$, $K_S^0 \pi^+\pi^-\pi^0$, and $K^+ K^-$.
The $\pi^0$ candidates used in this reconstruction are required to have momentum in the center-of-mass system greater than $100~{\rm MeV}/c$,
and energy in the laboratory system greater than $150~{\rm MeV}$.
If a $\pi^0$ is included, the reconstructed $D^0$ invariant mass is required to be within $16~{\rm MeV}/c^2$ of the nominal mass~\cite{PDG2020} corresponding to 85\% signal efficiency; otherwise, it is required to be within $8.5~{\rm MeV}/c^2$ corresponding to 91\% efficiency.
To improve the momentum resolution, a mass- and vertex-constrained fit is applied.
Candidates where the $\chi^2$ probability of the fit is less than 0.0001 are eliminated.\par

$\overline{D}{}^{*0}$ candidates are reconstructed in two decay modes: $\overline{D}{}^0\gamma$ and $\overline{D}{}^0\pi^0$.
For the $\overline{D}{}^0\gamma$ mode, only $\gamma$ candidates with an energy greater than $90~{\rm MeV}$ in the laboratory system are used.
For the $\overline{D}{}^0\pi^0$ mode, only $\pi^0$ candidates with a momentum in the center-of-mass system of less than $100~{\rm  MeV}/c$ and an energy in the laboratory system of less than $200~{\rm MeV}$ are used.
The difference in the reconstructed mass between $\overline{D}{}^{*0}$ and $\overline{D}{}^{0}$ is required to agree with the nominal value~\cite{PDG2020} within $9.0~{\rm MeV}/c^2$ and $2.0~{\rm MeV}/c^2$ for $\overline{D}{}^0\gamma$ and $\overline{D}{}^0\pi^0$, respectively,
corresponding to 90\% signal efficiency in each case.\par

$B$ meson candidates are then reconstructed in the decay modes $D^0\overline{D}{}^{*0} K^+$ and $D^0\overline{D}{}^{*0} K_S^0$.
To reduce wrong combinations, the daughter $K^+$ is required to have ${\cal L}_K/({\cal L}_\pi + {\cal L}_K) > 0.6$, corresponding to an identification efficiency of 89\%.
The $B$ candidates are selected based on the beam-energy constrained mass, $M_{\rm bc}\equiv \sqrt{(E_{\rm beam}^{\rm cms})^2-(p_{B}^{\rm cms})^2}$ and the difference of the energy in the center-of-mass system between the $B$ candidate and the beam, $\Delta E\equiv E_{B}^{\rm cms}-E_{\rm beam}^{\rm cms}$, where $E_{\rm beam}^{\rm cms}$ is the beam energy in the center-of-mass system corresponding to half of $\sqrt{s}$, and $p_{B}^{\rm cms}$ and $E_{B}^{\rm cms}$ are the energy and momentum of $B$ candidates in the center-of-mass system, respectively.
We retain events with $M_{\rm bc} > 5.2~{\rm GeV}/c^2$ and $|\Delta E| < 50~{\rm MeV}$ for later analysis. 
The $M_{\rm bc}$ signal region is defined as $|M_{\rm bc} - m_B| < 4.5~{\rm MeV}/c^2$ ($\approx 2\sigma$) for $\overline{D}{}^{*0} \to \overline{D}{}^0\gamma$ and $|M_{\rm bc} - m_B| < 6.0~{\rm MeV}/c^2$ ($\approx 2.5\sigma$) for $\overline{D}{}^{*0} \to \overline{D}{}^0\pi^0$, where $m_B$ denotes the nominal $B$ mass~\cite{PDG2020}.
The $\Delta E$ signal region is defined as $|\Delta E| < 12~{\rm MeV}$ ($\approx 2\sigma$).
For suppression of continuum events, we use a FastBDT classifier~\cite{fastBDTBelle} trained on the simulation sample with the following event-shape information as input: modified Fox-Wolfram moments~\cite{FW-and-KSFW}, the momentum flow in concentric cones around the thrust axis~\cite{cleoCone}, and thrust-related quantities.
Events for which the classifier output is less than 0.15 are eliminated.
This requirement retains 96\% of the signal candidates and rejects 49\% of the candidates of continuum events.
\par

\begin{figure*}[bt]
    \begin{minipage}{0.49\hsize}
	   \begin{center}
    	\includegraphics[width=\hsize]{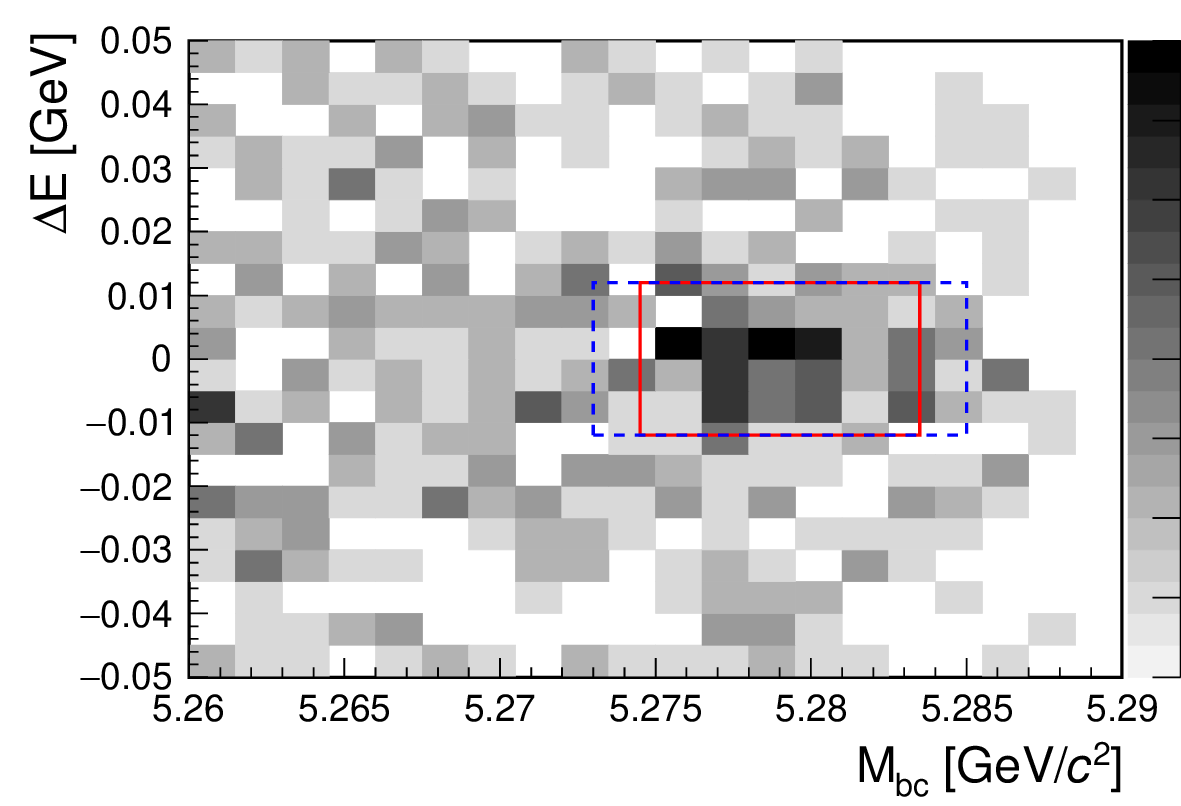}
    	\end{center}
    \end{minipage}
    \begin{minipage}{0.49\hsize}
        \begin{center} 
    	\includegraphics[width=\hsize]{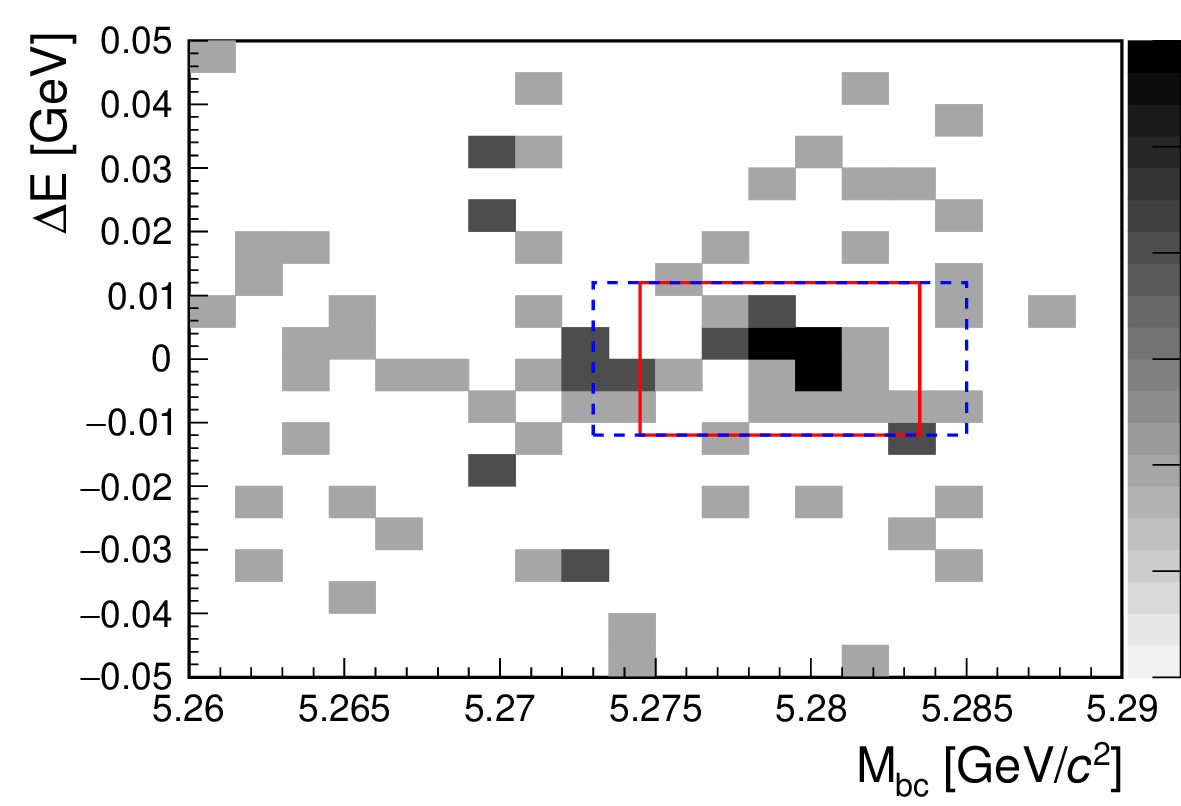}
 \end{center}
    \end{minipage}
    \caption{Distributions of $(M_{\rm bc},\Delta E)$ for $B^+$ (left) and $B^0$ (right) candidates in the $M(D^0\overline{D}{}^{*0}) < 3.88~{\rm GeV}/c^2$ region, where the signal-to-background ratio for $X(3872)$ in data is relatively high.
    The red solid and blue dashed rectangles show the $(M_{\rm bc},\Delta E)$ signal regions for $\overline{D}{}^{*0} \to \overline{D}{}^0\gamma$ and $\overline{D}{}^0 \pi^0$, respectively.
    \label{figID:deltaEVersusMbc}}
\end{figure*}

After this selection, the average number of $B$ candidates per event is 1.8, because $D^0\overline{D}{}^{*0}$ and $D^{*0}\overline{D}{}^{0}$ are often indistinguishable and double-counted.
To avoid multiple counting of signal events, we select the candidate that has the highest value of the product of the following likelihood ${\cal L}$ and prior probability ${\cal P}$

\begin{equation}
\begin{split}
{\cal L} =& {\cal L}_{M(D^0)} \times  {\cal L}_{M(\overline{D}{}^0)} \times {\cal L}_{M(\overline{D}{}^{*0}) - M(\overline{D}{}^0)}\\
&\qquad \qquad \qquad \qquad \qquad \times {\cal L}_{\Delta E}~[\times {\cal L}_{M(\pi^0)}],\\
{\cal P} =&
\frac{\varepsilon_{ijk}}{\zeta_{ijk}}\times {\cal B}(D^0 \to i) \times
{\cal B}(\overline{D}{}^0 \to j) \times {\cal B}(\overline{D}{}^{*0} \to k),
\end{split}
\end{equation}
where ${\cal L}$ is the product of the likelihoods of the measured $D^0$, $\overline{D}{}^0$, and $\overline{D}{}^{*0}$ masses, and $\Delta E$; 
and, for the $\overline{D}{}^{*0} \to \overline{D}{}^0\pi^0$ mode, the likelihood of the measured $\pi^0$ mass.
Each likelihood is obtained using probability density functions (PDFs) determined using the MC samples.
The probability ${\cal P}$ is obtained from the probability that a signal event can be reconstructed $\varepsilon_{ijk}$, the average number of $B$ candidates per event $\zeta_{ijk}$, and the decay branching fraction, when $D^0$, $\overline{D}{}^0$, and $\overline{D}{}^{*0}$ are reconstructed in the $i$, $j$, and $k$ modes, respectively.
The values of $\varepsilon_{ijk}$ and $\zeta_{ijk}$ are determined using the MC samples.
The $(M_{\rm bc},\Delta E)$ distribution of the selected $B$ candidates is shown in Fig.~\ref{figID:deltaEVersusMbc}.
The red solid (blue dashed) rectangle shows the $(M_{\rm bc},\Delta E)$ signal region for $\overline{D}{}^{*0} \to \overline{D}{}^0\gamma$ ($\overline{D}{}^0 \pi^0$): 
$B$ candidates used in the lineshape study are selected from this region.
\par

For all events remaining in the selection, the following $D^0\overline{D}{}^{*0}$ invariant mass is calculated instead of applying a mass-constrained fit to improve the mass resolution,
\begin{equation}
    \label{analysis:recoAndSelection:eqID:definitionXMass}
    \begin{split}
    M&(D^0\overline{D}{}^{*0}) \\
    &= \left\{
        \begin{array}{ll}
        M(D^0\overline{D}{}^0\gamma) - M(\overline{D}{}^0\gamma) + m_{\overline{D}{}^{*0}} &\\ 
        ~~~~~~~~~~~~~~~~~~~~~~~~~~~~~{\rm for~}\overline{D}{}^{*0} \to \overline{D}{}^0 \gamma, \\
        M(D^0\overline{D}{}^0\pi^0) - M(\overline{D}{}^0\pi^0) + m_{\overline{D}{}^{*0}} &\\
        ~~~~~~~~~~~~~~~~~~~~~~~~~~~~~{\rm for~}\overline{D}{}^{*0} \to \overline{D}{}^0 \pi^0, \\
        \end{array}
    \right.
    \end{split}
\end{equation}
where the reconstructed $\overline{D}{}^{*0}$ invariant mass, $M(\overline{D}{}^0\gamma)$ or $M(\overline{D}{}^0\pi^0)$, is subtracted, and the $\overline{D}{}^{*0}$ nominal mass, $m_{\overline{D}{}^{*0}}$, is added.
The lineshape and signal yield are determined by fitting the distribution in the region below $4.0~{\rm GeV}/c^2$.

\section{Fit Strategy and Detector Response}
\label{secID:fitmodel}
In this work, the obtained $M(D^0 \overline{D}{}^{*0})$ distributions are fitted with two lineshape models: the relativistic Breit-Wigner, and a Flatt\'{e}-inspired model.
The fits with these two models, shown in the next section, are performed with the following procedure.\par

When signal events are reconstructed correctly, the invariant mass distribution has a peak consisting of the natural lineshape convolved with the mass-dependent detector response.
This response, \emph{i.e.}\ the mass dependence of the signal efficiency and the mass resolution, is studied and parameterized using a set of $X(3872) \to D^0 \overline{D}{}^{*0}$ MC samples generated with zero width, and a range of mass values from the $D^0\overline{D}{}^{*0}$ threshold to $4.0~{\rm GeV}/c^2$.
Here, the $X(3872) \to D^0 \overline{D}{}^{*0}$ decays are generated using a uniform phase space model; the $D^{*0}$ width is assumed to be around 60~keV~\cite{Braaten2007}.
Since the signal-to-noise ratio depends on the $D^{*0}$ decay mode, fits are performed separately for $D^{*0} \to D^0 \gamma$ and $D^{*0} \to D^0 \pi^0$.
In addition, fits are performed separately for $B^0$ and $B^+$ candidates to determine the ratio of branching fractions between $B^0 \to X(3872) K^0$ and $B^+ \to X(3872) K^+$.\par

\begin{figure*}[bt]
    \begin{minipage}{0.49\hsize}
	   \begin{center}
    	\includegraphics[width=\hsize]{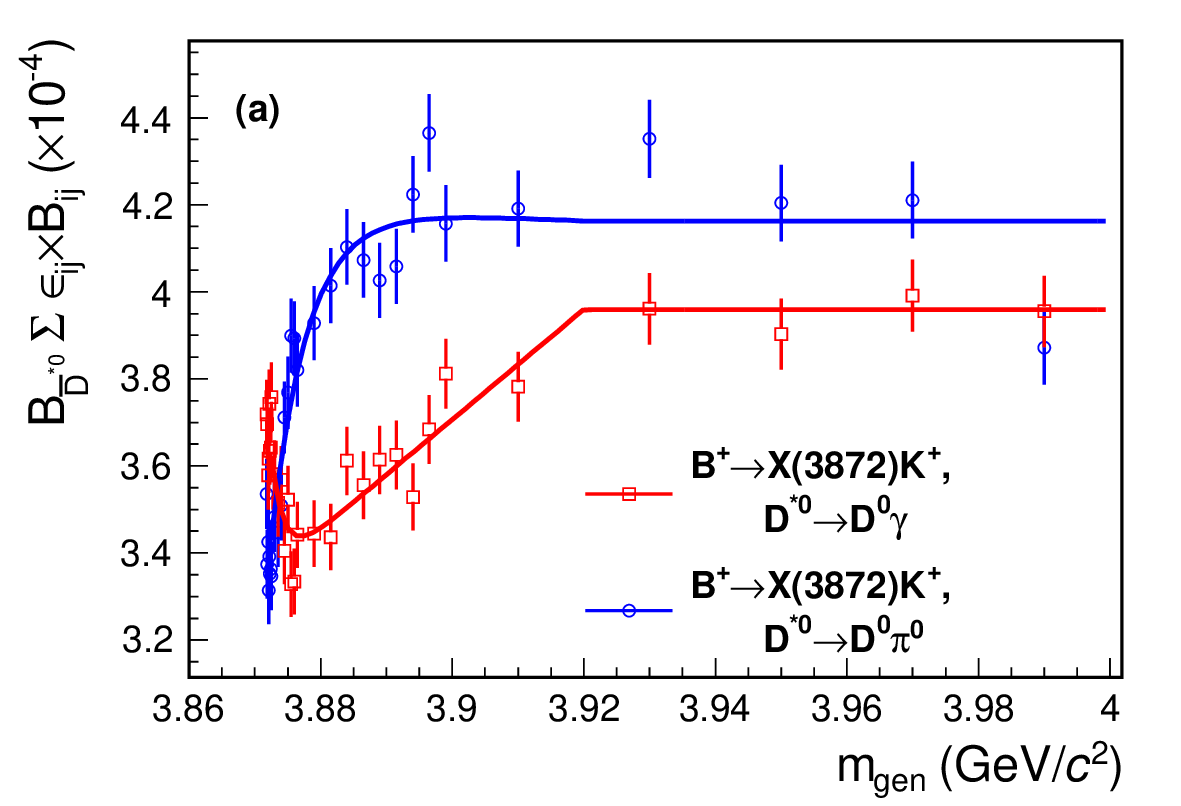}
    	\end{center}
    \end{minipage}
    \begin{minipage}{0.49\hsize}
        \begin{center} 
    	\includegraphics[width=\hsize]{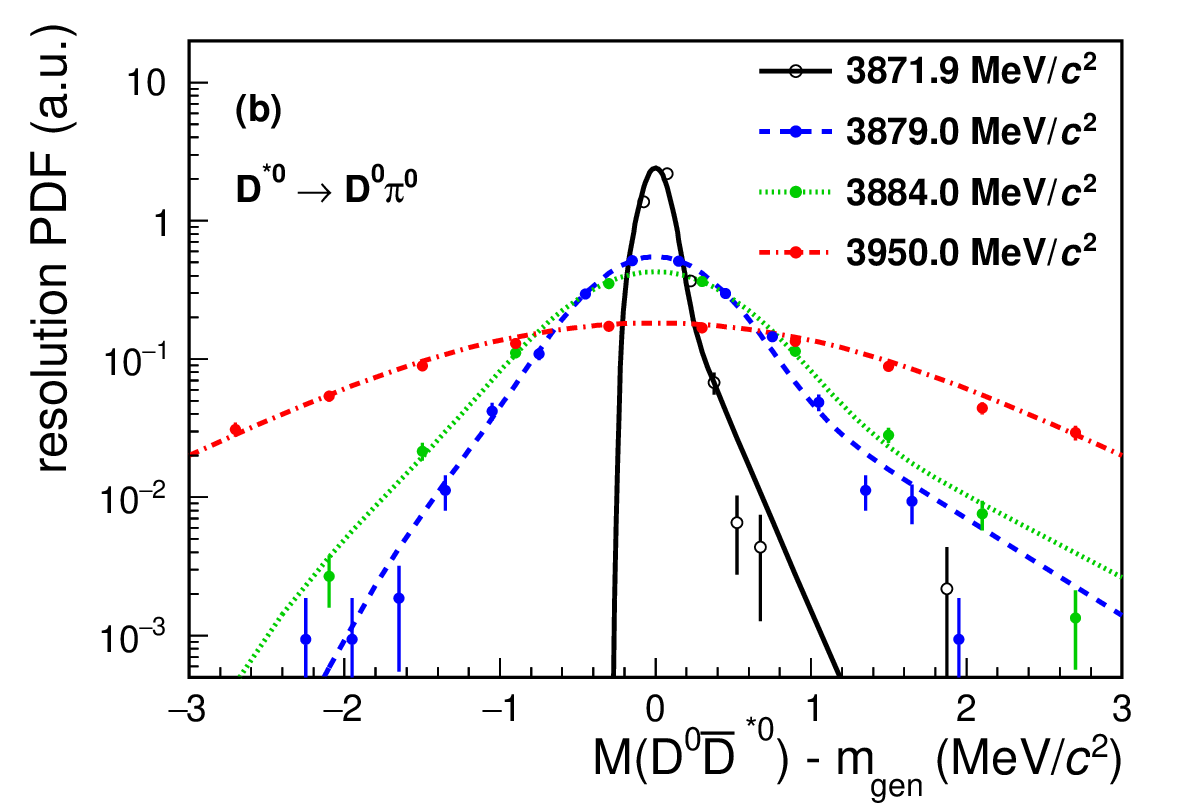}
    	\end{center}
    \end{minipage}
     \caption{
     The detector response for the signal component.
     (a) The sum of products of the signal efficiency and the branching fraction of the intermediate states ${\cal B}_{\overline{D}{}^{*0}} \sum \epsilon_{ij} \times {\cal B}_{ij}$ as a function of the $X(3872)$ mass generated in the MC samples for $B^+\to X(3872)K^+$; the blue circles and the red squares are for $D^{*0}\to D^0\pi^0$ and $D^{*0}\to D^0\gamma$, respectively.
     The lines represent the parameterized efficiency functions. For $B^0\to X(3872)K^0$, similar structures are obtained with a ratio of $B^+\to X(3872)K^+$ to $B^0\to X(3872)K^0$ of almost $4:1$.
     (b) The $M(D^0\overline{D}{}^{*0})$ spread due to the detector response for the $X(3872)$ lineshape generated with zero width and masses of $3871.9~{\rm MeV}/c^2$, $3879.0~{\rm MeV}/c^2$, $3884.0~{\rm MeV}/c^2$, and $3950.0~{\rm MeV}/c^2$ for the $D^{*0}\to D^0\pi^0$ decay mode. The circles show normalized distributions obtained from the MC sample. The curves show the parameterized resolution functions. Similar results are obtained for $D^{*0}\to D^0\gamma$.
     \label{figID:detectorResponse}}
\end{figure*}

The fit function for correctly reconstructed $X(3872) \to D^0 \overline{D}{}^{*0}$ decays, which we refer to as ``signal'', is constructed as follows.
The signal efficiency varies depending on the mass by a few tens of percent, especially around the threshold, as shown in Fig.~\ref{figID:detectorResponse}~(a).
It is parameterized by the threshold function $p_0\{1-p_1e^{p_2(M-m_{D^0}-m_{D^{*0}})} + p_3(M-m_{D^0}-m_{D^{*0}})\}$ with parameters $p_0$--$p_3$ in the low-mass region, which is continuously connected to a constant value in the high-mass region.
The mass resolution for the signal is modeled as the sum of a Gaussian and a reversed Crystal Ball function~\cite{crystalball} with a common mean.
Figure~\ref{figID:detectorResponse}~(b) shows the $M(D^0\overline{D}{}^{*0})$ spread due to the resolution, 
and the resolution function used in this work, for several choices of the $X(3872)$ mass.
As noted in the previous Belle study, the resolution degrades with the square root of the difference between the mass and the threshold~\cite{Aushev2010}.
The convolution with the mass-dependent resolution function entails longer computation times.
The effect of smearing due to the resolution is small at masses away from the peak, since the natural lineshape is broad~\cite{BaBarDDstar, Aushev2010}.
For example, the full width at half maximum (FWHM) of the natural lineshape is a few MeV, while the FWHM of the mass resolution near the peak is only about 220 keV.
Therefore, instead of convolution with the mass-dependent resolution function, convolution with the specific resolution function at the mass of $3871.9~{\rm MeV}/c^2$ (near the peak) is adopted as an approximation.
To reproduce the behavior near threshold, the signal function is multiplied by a soft threshold function that rises from zero to one at the threshold using an error function.
The procedure is validated on $X(3872) \to D^0 \overline{D}{}^{*0}$ MC samples generated with a broad lineshape.
The effect of the approximation is negligible.
\par

The ratios of signal yields among the decay modes are fixed in the fit using the product of the expected total signal efficiency and the branching fraction of each decay mode.
Here the expected signal efficiency depends on the lineshape because of the mass dependence of the signal efficiency.
The total signal efficiency is obtained by averaging the signal efficiencies as a function of mass weighted by the values of the lineshape function.
It is then corrected by taking account of the signal which may leak out of the fit range, depending on the lineshape, and by taking the ratio of the area of the signal function in the fit range to that from the threshold to $m_B - m_K$.
Here, a mass-dependent resolution function is convolved with the signal function, because smearing due to the resolution is important at higher masses.
The calculation of the total signal efficiency is validated on MC samples for a broad range of lineshape parameters.\par

A separate fit function is used for ``broken signal'': cases where the $D^0$ is reconstructed incorrectly, a wrong $\pi^0$ or $\gamma$ is combined in the $\overline{D}{}^{*0}$ reconstruction, or a ${D}^{*0} \overline{D}{}^0$ signal event is misinterpreted as $D^0 \overline{D}{}^{*0}$ by combining $\pi^0$ or $\gamma$ from ${D}^{*0}$ incorrectly with the $\overline{D}{}^0$ to make a fake $\overline{D}{}^{*0}$.  
For the $\overline{D}{}^{*0} \to \overline{D}{}^0\pi^0$ mode, such events produce a broad peak in the $M(D^0 \overline{D}{}^{*0})$ signal region and possibly distort the lineshape of the signal. 
The fit function for the broken signal therefore takes account of the mass dependencies of the resolution and the efficiency as in the case of correctly reconstructed signal events.
The mass dependence of the efficiency is parameterized using the same threshold function used for the signal. 
The resolution is reproduced by a triple Gaussian multiplied by a soft threshold function at the $D^0\overline{D}{}^{*0}$ threshold, and its mass dependence is studied and parameterized using zero-width MC samples.
Since the resolution for the broken signal is several times worse than that for the signal, we do not use approximated convolution: we instead use a discrete convolution with the mass-dependent resolution function.
In the fit, the yield of the broken signal relative to the signal is fixed to the value expected from the lineshape and the ratio of the total efficiency of the broken signal to that of the signal.\par

The broken signal peak due to the $\overline{D}{}^{*0} \to \overline{D}{}^0\gamma$ mode has little sensitivity to the natural lineshape.
To reduce the systematic uncertainty due to the shape, we use a histogram PDF depending on the lineshape.
This PDF is obtained by plotting the broken-signal histogram for each of the zero-width MC samples, scaling it by the value of the assumed lineshape at the generated mass, and summing up all of the scaled histograms.
Here, the bin widths are adjusted to increase as the mass increases to suppress statistical fluctuations.

The background from $e^+e^- \to q\bar{q}$ ($q = u, d, s, c$) continuum events, and $e^+e^- \to \Upsilon(4S)\to B\overline{B}$ events other than signal, is studied using the background MC sample.
The shape of the invariant mass distribution is reproduced using a threshold function, $\sqrt{M-(m_{D^0}+m_{D^{*0}})}$, where $m_{D^0}$ and $m_{D^{*0}}$ are the nominal masses of $D^0$ and $D^{*0}$~\cite{PDG2020}, respectively.

\begin{figure*}[bt]
\includegraphics[width=\hsize]{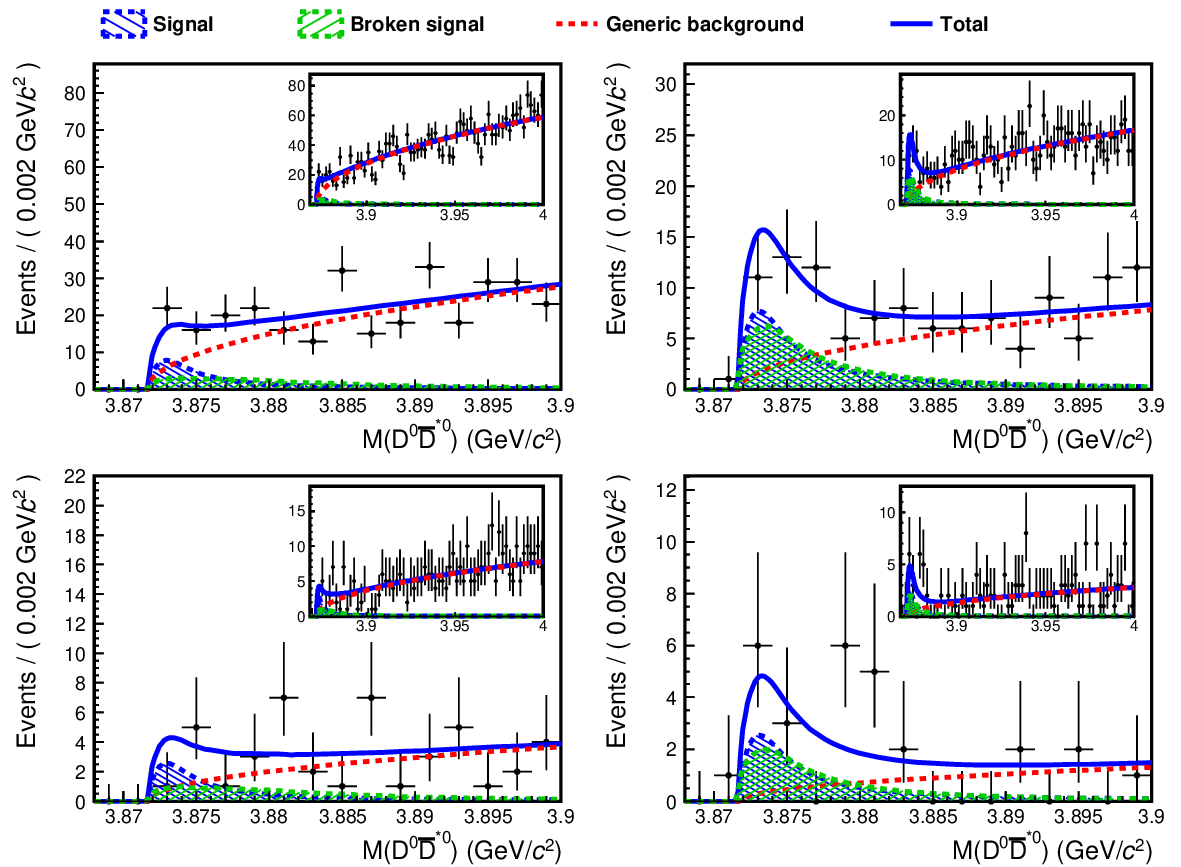}
\caption{The $M(D^0\overline{D}{}^{*0})$ distributions with the fit result with the relativistic Breit-Wigner lineshape for $B^+ \to X(3872)K^+$ (top) and $B^0 \to X(3872)K^0$ (bottom).
The left and right rows are for $\overline{D}{}^{*0}\to \overline{D}{}^0 \gamma$ and $\overline{D}{}^{*0}\to \overline{D}{}^0 \pi^0$, respectively. The points with error bars represent data. The blue solid line shows the total fit result. The blue and green dashed lines show the signal contributions and broken signal contributions, respectively.
The red dashed line shows the contribution of generic background.
 \label{figID:fitBW}}
\end{figure*}

\section{Fit to Data}
\label{secID:fitresult}
\subsection{Relativistic Breit-Wigner model}
The relativistic Breit-Wigner lineshape function is defined as~\cite{PDG2020}
\begin{equation}
\label{introduction:lineshapemodel:BreitWigner:eq:definition}
    f_{\rm BW}(M) = \frac{m_{\rm BW} M \Gamma(M)}{(M^2-m_{\rm BW}^2)^2 + m_{\rm BW}^2\Gamma(M)^2},
\end{equation}
where $M$ is the observed invariant mass, and $m_{\rm BW}$ is the mass of the resonance. 
The mass-dependent width $\Gamma(M)$ is defined as 
\begin{equation}
    \Gamma(M)=\Gamma_{\rm BW} \frac{m_{\rm BW}}{M}\biggr(\frac{p(M)}{p(m_{\rm BW})}\biggl)^{2L+1},
\end{equation}
where $\Gamma_{\rm BW}$ and $L$ are the width of the resonance and the orbital angular momentum, respectively.
Taking account of the closeness to the threshold, the decay is assumed to be pure S-wave ($L = 0$) with no D-wave ($L=2$) admixture.
The momentum of one of the daughters in the rest frame of $X(3872)$, $p(M)$, can be calculated as
\begin{equation}
\begin{split}
    p(M) =& \frac{1}{2M} \sqrt{(M^2 - (m_{D^0}+m_{D^{*0}})^2)}\\
    &\times \sqrt{(M^2 - (m_{D^0}-m_{D^{*0}})^2)}.
\end{split}
\end{equation}
\par

Figure~\ref{figID:fitBW} presents the $M(D^0\overline{D}{}^{*0})$ distributions obtained from the data.
Here, unbinned maximum likelihood fits are performed simultaneously to the distributions for the $\overline{D}{}^{*0}$ decay modes, $\overline{D}{}^{*0} \to \overline{D}{}\pi^0$ and $\overline{D}{} \gamma$, and for the $B^+$ and $B^0$ samples, with common fit parameters $m$ and $\Gamma_0$.
The PDFs are convolved with the detector response as described in the previous section.
Table~\ref{tbID:fitBW} summarizes the parameters obtained from the fit.
The significance is determined from the log-likelihood ratio $-2\ln({\cal L}_0/{\cal L})$ accounting for the difference in the number of degrees of freedom, where ${\cal L}_0$ and ${\cal L}$ are the fit likelihood without and with the peak component; \emph{i.e.}, the yield is constrained to be zero for the significance of each $B$ mode, and the parameters $m$ and $\Gamma_0$ are additionally dropped for the combined significance.
Here the likelihood is smeared to take account of the systematic uncertainties on the signal yields as described below.
The significance is found to be $5.9\sigma$ for $B^+ \to X(3872)K^+$, and $5.2\sigma$ for $B^0 \to X(3872) K^0$.
The absence of peaks in the $M(D^0\overline{D}{}^{*0})$ distribution in the $(M_{\rm bc},\Delta E)$ sideband region confirms that any contribution from peaking background is small; here the sideband region is defined as $12~{\rm MeV}/c^2 < |M_{\rm bc}-m_B| < 20~{\rm MeV}/c^2 $ or $30~{\rm MeV} < |\Delta E| < 50~{\rm MeV}$.

\begin{table*}[tb]
\caption{Results using the relativistic Breit-Wigner lineshape: the fitted mass, width and signal yield, the total signal efficiency, and the significance.}
\label{tbID:fitBW}
{\renewcommand{\arraystretch}{1.2}
\begin{tabular}{c@{\hspace{2ex}}c@{\hspace{2ex}}c@{\hspace{2ex}}c@{\hspace{2ex}}c@{\hspace{2ex}}c} \hline \hline
Mode &  $m~({\rm MeV}/c^2)$ & $\Gamma_0~({\rm MeV})$ & $N_{\rm sig}$ & ${\cal B}_{\overline{D}{}^{*0}} \sum \epsilon_{ij} \times {\cal B}_{ij}$ & Significance\\
\hline
Combined & $3873.71^{+0.56}_{-0.50}$ & $5.2^{+2.2}_{-1.5}$ & $70.5^{+13.6}_{-11.5}$ & $8.70 \times 10^{-4}$ & $7.5\sigma$\\
$X(3872) K^+$ & --- & --- & $53.2^{+11.6}_{-\phantom{1}9.8}$ & $6.92 \times 10^{-4}$ & $5.9\sigma$\\
$X(3872) K^0$ & --- & --- & $17.3^{+\phantom{1}4.7}_{-\phantom{1}4.1}$ & $1.78 \times 10^{-4}$ & $5.2\sigma$\\
\hline \hline
\end{tabular}
}
\end{table*}

\begin{table*}[bt]
    \centering
    \caption{Summary of systematic uncertainty for the mass, width, and branching fractions measurements using the relativistic Breit-Wigner lineshape. \label{tabID:systematicBW}}
    \setlength{\tabcolsep}{1pt}
    \begin{tabular}{r l  c  c  c  c  c}\hline  \hline
        \multicolumn{2}{c}{Source} & $m~({\rm MeV}/c^2)$ & $\Gamma_0~({\rm MeV})$ & $X(3872)K^+$~(\%)& $X(3872)K^0$~(\%) & Ratio($K^0/K^+$)~(\%) \\ \hline
        (i) & Generic BG PDF & $\pm0.07$ & $\pm0.38$ & $\pm\phantom{1}8.2$ & $\pm1.4$ & $\pm6.7$ \\
        (ii) & Mass resolution & $\pm0.02$ & $-0.11$/$+0.13$ & $-0.2$/$+0.4$ & $-0.3$/$+0.4$ & $-0.1$/$+0.0$ \\
        (iii) & Mass dependence of efficiency  & $\pm0.02$ & $-0.08$/$+0.07$  & $-2.7$/$+2.0$ & $-2.3$/$+1.7$ & $-0.5$/$+0.6$\\
        (iv) & Ratio of broken-signal BG to signal & $\pm0.01$ & $\pm0.02$  & $\pm\phantom{1}2.1$ & $\pm0.6$ & $\pm2.1$\\
        (v) & Fit bias & $-0.02$/$+0.00$ & $-0.02$/$+0.00$ & $-1.3$/$+0.0$ & $-7.3$/$+0.0$ & $-4.5$/$+0.0$\\
        (vi) & $D^{*0}$ and $D^0$ masses & $\pm 0.10$ & $\cdots$ & $\cdots$ & $\cdots$ & $\cdots$\\ 
        (vii) & $D^{*0}$ width & $-0.01$/$+0.02$ & $\pm0.02$ &  $\pm\phantom{1}0.0$ & $\pm0.0$ & $\pm0.0$ \\ 
        (viii) & Broken-signal shape for $\overline{D}{}^{*0}\to \overline{D}{}^0\gamma$ & $\pm 0.00$ & $\pm 0.01$ & $\pm \phantom{1}0.1$ &$\pm 0.1$ & $\pm 0.1$\\
        (ix) & Signal ratio of $\overline{D}{}^{*0}\to \overline{D}{}^0\gamma$ to $\overline{D}{}^0\pi^0$ & $\pm 0.01$ & $\pm 0.05$ & $\pm\phantom{1}0.8$ & $\pm 0.2$ & $\pm 0.6$ \\\hline 
        (x) & Tracking efficiency & $\cdots$ & $\cdots$ & $\pm\phantom{1}2.1$ & $\pm2.4$ & $\pm0.3$ \\
        (xi) & PID efficiency & $\cdots$ & $\cdots$ & $\pm\phantom{1}2.9$ & $\pm2.4$ & $\pm0.4$\\
        (xii) & $K_S^0$ efficiency & $\cdots$ & $\cdots$ & $\pm\phantom{1}0.2$ & $\pm1.0$ & $\pm0.8$\\
        (xiii) & $\pi^0$ reconstruction & $\cdots$ & $\cdots$ & $\pm\phantom{1}1.9$ & $\pm1.9$ & $\cdots$\\
        (xiv) & $\gamma$ reconstruction & $\cdots$ & $\cdots$ & $\pm\phantom{1}1.5$ & $\pm1.5$ & $\cdots$\\
        (xv) & $\sum \epsilon_{ij} \times {\cal B}_{ij}$ & $\cdots$ & $\cdots$ & $\pm\phantom{1}1.4$ &  $-3.1$/$+2.3$ & $-1.7$/$+0.9$\\
        (xvi) & $N_{B\overline{B}}$ & $\cdots$ & $\cdots$ & $\pm\phantom{1}1.4$ & $\pm1.4$ & $\cdots$ \\
        (xvii) & ${\cal B} (\Upsilon(4S) \to B\overline{B})$ & $\cdots$ & $\cdots$ & $\pm\phantom{1}1.2$ & $\pm1.2$ &$\pm2.4$\\ \hline
        & Total & $\pm0.13$ & $\pm0.4\phantom{0}$ &  $\pm 10\phantom{.0}$ & $-9.6$/$+5.7$ & $-9.0$/$+7.6$\\ \hline \hline
    \end{tabular}
\end{table*}

The lineshape parameters are determined to be
\begin{equation*}
\begin{split}
m_{\rm BW} &= 3873.71 ^{+0.56}_{-0.50} ({\rm stat}) \pm 0.13 ({\rm syst}) ~{\rm MeV}/c^2,\\
\Gamma_{\rm BW} &= 5.2 ^{+2.2}_{-1.5} ({\rm stat}) \pm 0.4 ({\rm syst})~{\rm MeV}.
\end{split}
\end{equation*}
The difference between $m_{\rm BW}$ and the $D^0\overline{D}{}^{*0}$ threshold is found to be
\begin{equation*}
\begin{split}
m_{\rm BW} - &(m_{D^0}+m_{D^{*0}}) \\
&= 2.02^{+0.56}_{-0.50} ({\rm stat}) \pm 0.08 ({\rm syst})~{\rm MeV}/c^2.
\end{split}
\end{equation*}
\par

The systematic uncertainties are listed in Table~\ref{tabID:systematicBW}.
We consider the following nine sources of uncertainty on the mass, the width, and the signal yield:
(i) The uncertainty due to the assumed shape of the generic background is estimated by performing a fit after changing the PDF from the threshold function with a square root to an inverted ARGUS function~\cite{ARGUSfunc}.
(ii) The mass resolution is validated by comparing the data and MC $\Delta E$ resolution in the $B^+ \to \overline{D}{}^{*0} \pi^+\pi^-\pi^+$ control sample, which has a similar decay topology to $B \to X(3872)(\to D^0 \overline{D}{}^{*0}) K$. 
The ratios of the mass resolution obtained for MC and data are $1.01 \pm 0.10$ for $\overline{D}{}^{*0}\to \overline{D}{}^0\gamma$ and $1.08 \pm 0.13$ for $\overline{D}{}^{*0}\to \overline{D}{}^0\pi^0$.
This resolution is consistent in data and MC, so no correction is applied, and the associated uncertainty is assigned by performing fits with the resolution varied by the precision, $\pm 1\sigma \equiv \pm 13\%$.
(iii, iv) The uncertainties arising from the mass dependence of the efficiency and the ratio of the broken-signal to the signal are evaluated by summing in quadrature the changes induced by $\pm 1\sigma$ variations of the relevant parameters.
(v) The uncertainty due to possible bias in the fit is evaluated by performing pseudo experiments.
The input value of a parameter subtracted from the median of the parameter distribution is regarded as the corresponding uncertainty.
(vi) For the $m_{\rm BW}$ measurement only, the uncertainty arising from the finite precision of the $D^0$ mass and the $\Delta(m_{D^{*0}}-m_{D^0})$ mass difference is taken as the $\pm 1\sigma$ uncertainty of $2m_{D^0} + \Delta(m_{D^{*0}}-m_{D^0}) = 3871.69 \pm 0.10~{\rm MeV}/c^2$ following Ref.~\cite{PDG2020}.
(vii) The nonzero $D^{*0}$ width ($\Gamma_{D^{*0}}$) leads to three potential sources of bias: a bias arising from the mass difference technique, a bias arising from the consideration of the $D^{*0}$ width in the lineshape model, and a bias due to the interference between $X(3872) \to D^0\overline{D}{}^{*0}$ and $\overline{D}{}^0 D^{*0}$.
Biases from these three sources are evaluated as follows.
For the first bias, two $M(D^0\overline{D}{}^{*0})$ distributions are formed in MC with a broad lineshape: one where $m_{D^{*0}}$ in Eq.~\eqref{analysis:recoAndSelection:eqID:definitionXMass} is fixed to the nominal value (as in our analysis), and the other where $m_{D^{*0}}$ is replaced 
by the true $D^{*0}$ mass generated by \texttt{EvtGen}, where $\Gamma_{D^{*0}} = 65.5~{\rm keV}$~\cite{Braaten2007} is assumed.
Each distribution is fitted with the PDF of the signal component, and the largest difference is regarded as the associated uncertainty.
For the second bias, the distribution for data is fitted after smearing the assumed lineshape with a Breit-Wigner function of $\Gamma_{D^{*0}} = 65.5~{\rm keV}$, and the change from the original result is regarded as the associated uncertainty.
The third bias is ignored since the interference effect is negligible above the threshold~\cite{Hanhart2010}.
The uncertainties associated with the first and second biases are added in quadrature.
(viii) Limited MC statistics lead to uncertainty on the shape of the broken-signal for $\overline{D}{}^{*0}\to \overline{D}{}^0\gamma$.
This is evaluated by repeating the fit while varying each bin entry of the MC PDF histogram assuming Poisson distributions.
The 68\% interval of the distributions of the resulting fit values is used to assign the uncertainty.
(ix) The uncertainty arising from the fixed ratio of the signal yields for $\overline{D}{}^{*0}\to \overline{D}{}^0\gamma$ to $\overline{D}{}^0\pi^0$ is evaluated by performing new fits, and varying the ratio of branching fractions between $D^{*0}\to D^0\gamma$ and $D^{*0}\to D^0\pi^0$ by $\pm1\sigma$~\cite{PDG2020}.
The difference from the original result is treated as the uncertainty.\par

The product branching fraction is calculated as follows:
\begin{equation}
\label{fitting:flatte:eq:branchingFractionCalc}
\begin{split}
    {\cal B} & (B\to X(3872)K) \times {\cal{B}}(X(3872) \to D^0\overline{D}{}^{*0}) \\
    & \qquad = \frac{N_{\rm sig}}{2N_{B\overline{B}} {\cal B}(\Upsilon(4S) \to B\overline{B}) {\cal B}_{\overline{D}{}^{*0}} \sum \epsilon_{ij} \times {\cal B}_{ij}}, \\
\end{split}
\end{equation}
where ${\cal B}_{\overline{D}{}^*}$ is the appropriate $\overline{D}{}^{*0}$ branching fraction, and $\sum \epsilon_{ij} \times {\cal B}_{ij}$ is the sum of efficiencies multiplied by the product of branching fractions for the various $D^0 \to i$ and $\overline{D}{}^0 \to j$ decay modes used.
For ${\cal B}(\Upsilon(4S)\to B\overline{B})$, 0.514 and 0.486 are assigned for the $B^+B^-$ and $B^0\overline{B}{}^0$ modes, respectively~\cite{PDG2020}.
The results are
\begin{equation*}
\begin{split}
    {\cal B}& (B^+\to X(3872)K^+) \times {\cal{B}}(X(3872) \to D^0\overline{D}{}^{*0}) \\
    &  \qquad = (0.97 ^{+0.21}_{-0.18} ({\rm stat}) \pm 0.10 ({\rm syst})) \times 10^{-4},\\
    {\cal B}&(B^0 \to X(3872)K^0) \times {\cal{B}}(X(3872) \to D^0\overline{D}{}^{*0}) \\
    & \qquad = (1.30 ^{+0.36}_{-0.31} ({\rm stat}) ^{+0.12}_{-0.07} ({\rm syst})) \times 10^{-4}.\\
\end{split}
\end{equation*}

Here we consider the following eight sources of systematic uncertainties in addition to those previously described for the lineshape parameters;
(x) The uncertainty of the tracking efficiency is estimated using a $D^{*+} \to \pi^+ D^0 (\to \pi^+ \pi^- K_S^0)$ sample for tracks with high momentum.
The efficiency is consistent in data and MC; the precision of the test, 0.35\% per track, is taken as a systematic uncertainty.
For tracks with low momentum, the sample of soft $\pi^-$ in $D^{*-} \to \overline{D}^0 \pi^-$ in the $B^0 \to D^{*-} \pi^+$ decay is used. The ratio of tracking efficiency obtained for MC and data is applied as a correction factor. The uncertainty in the correction factor is regarded as a systematic uncertainty.
(xi, xii, xiii) Efficiencies for hadron identification, $K_S^0$ selection, and $\pi^0$ detection are evaluated using control samples:
$D^{*+} \to D^0 (\to K^-\pi^+)\pi^+$, $D^{*+} \to D^0 (\to K_S^0\pi^0)\pi^+$, and $\tau^- \to \pi^-\pi^0 \nu_\tau$, respectively.
In each case a correction factor is applied to the signal efficiency based on the ratio of efficiencies obtained for MC and data,
and the uncertainty on the correction factor is taken as the associated systematic uncertainty. 
(xiv) The uncertainty of the efficiency of $\gamma$ detection is evaluated using a $B^+ \to \chi_{c1} (\to J/\psi\gamma) K^+$ sample: 3.0\% is assigned for the $D^{*0} \to D^0\gamma$ decay mode.
(xv) The uncertainty on $\sum \epsilon_{ij} \times {\cal B}_{ij}$ mainly arises from the uncertainties on the $D^0$ branching fractions, and the limited size of the signal MC sample.
In addition, validation of the calculation method for the total signal efficiency shows input-output differences in the $B^0$ decay mode larger than expected
 from statistical fluctuations: the largest of these is assigned as a systematic uncertainty. 
The uncertainties from these sources are added.
(xvi) The number of $B\overline{B}$ pairs in the data set is measured to be $(772 \pm 11) \times 10^6$: the associated uncertainty is set to 1.4\%.
(xvii) The uncertainties on the branching fractions ${\cal B}(\Upsilon(4S) \to B^+B^-) = (51.4\pm 0.6)\%$ and ${\cal B}(\Upsilon(4S) \to B^0\overline{B}{}^0) = (48.6\pm 0.6)\%$~\cite{PDG2020} are also included.
\par

The ratio of branching fractions between $B^0 \to X(3872) K^0$ and $B^+ \to X(3872) K^+$ is found to be 
\begin{equation*}
    \frac{{\cal B}(B^0\to X(3872)K^0)}{{\cal B}(B^+ \to X(3872)K^+)} = 1.34^{+0.47}_{-0.40} ({\rm stat}) ^{+0.10}_{-0.12} ({\rm syst}),
\end{equation*}
with the same sources of systematic uncertainty as for the branching fractions; some sources cancel, or partially cancel, in the ratio (see Table~\ref{tabID:systematicBW}).

\subsection{Flatt\'{e} model}
\begin{figure*}[tb]
\includegraphics[width=\hsize]{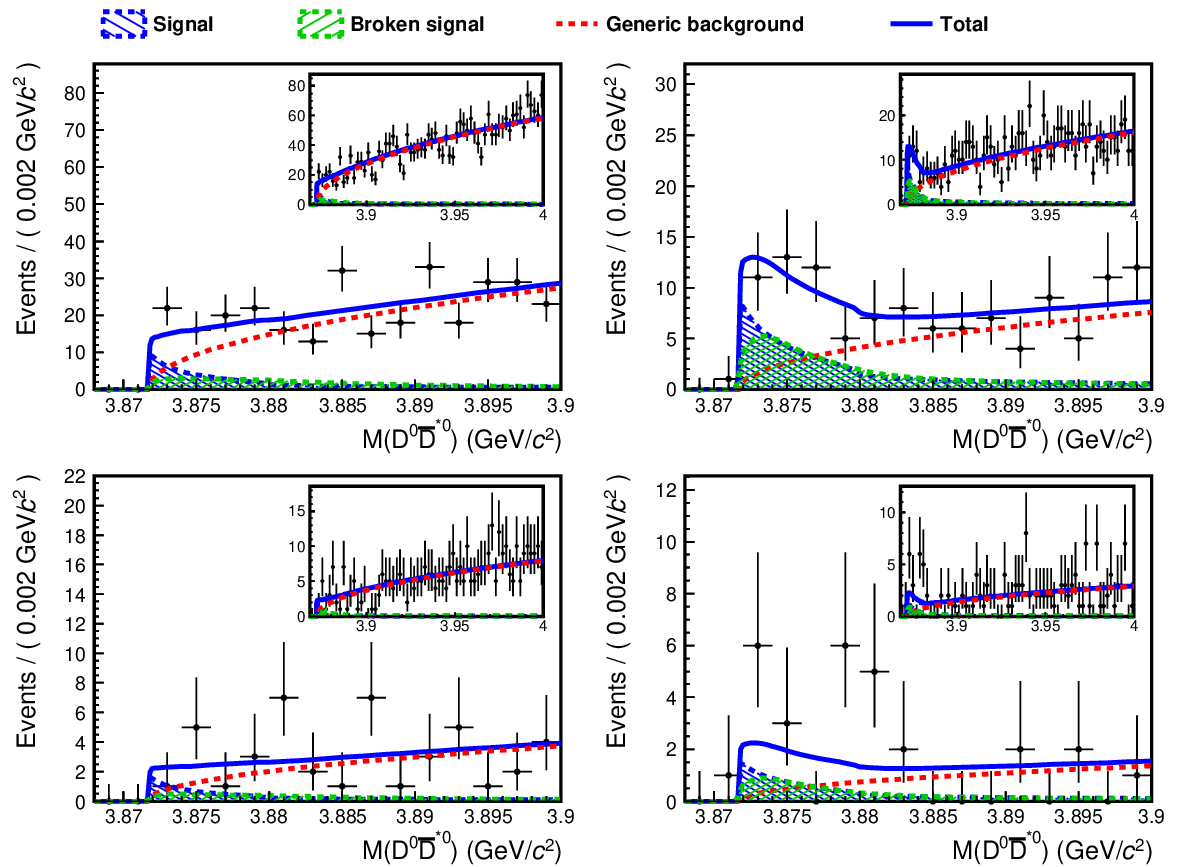}
\caption{The $M(D^0\overline{D}{}^{*0})$ distributions with the fit result with the \Flatte lineshape for $B^+ \to X(3872)K^+$ (top) and $B^0 \to X(3872)K^0$ (bottom).
The left and right rows are for $\overline{D}{}^{*0}\to \overline{D}{}^0 \gamma$ and $\overline{D}{}^{*0}\to \overline{D}{}^0 \pi^0$, respectively. The points with error bars represent data. The blue solid line shows the total fit result. The blue and green dashed lines show the signal contributions and broken-signal contributions, respectively.
The red dashed line shows the contribution of generic background.}
\label{figID:fitFlatte}
\end{figure*}

The Flatt\'{e}-inspired parametrization is defined as follows using the energy from the $D^0\overline{D}{}^{*0}$ threshold, $E = M - (m_{D^0}+m_{D^{*0}})$~\cite{Hanhart2007, Kalashnikova2009, unitInSecVB}:
\begin{equation}
    f_{\rm Flatte}(E) = \frac{gk_{D^0\overline{D}{}^{*0}}}{|D(E)|^2},
\end{equation}
\begin{equation}
D(E) = \left\{ 
    \begin{array}{ll}
    E - E_f -\frac{1}{2}g\kappa_{{D^+D^{*-}}}+ \frac{i}{2}[gk_{D^0\overline{D}{}^{*0}} + \Gamma(E)] &\\
     ~~~~~~~~~~~~~~~~~~~~~~~~~~~~~~~~~~~{\rm for}~0< E < \delta,\\
    E - E_f + \frac{i}{2}[g(k_{D^0\overline{D}{}^{*0}}+k_{{D^+D^{*-}}}) + \Gamma(E)] &\\
     ~~~~~~~~~~~~~~~~~~~~~~~~~~~~~~~~~~~{\rm for}~E > \delta,
    \end{array}
\right.
\end{equation}
where $E_f = m_0 - (m_{D^0} + m_{D^{*0}})$ is the mass difference of this state ($m_0$) from the threshold, and $g$ is the coupling constant for the $D\overline{D}{}^{*}$ channels;
we assume the coupling constants for the $D^0\overline{D}{}^{*0}$ and $D^+D^{*-}$ channels are the same due to isospin symmetry.
The momenta $k_a$ and $\kappa_a$ for the channel $a$ are measured in the rest frame of the $X(3872)$.
They are expressed using the reduced mass $\mu$ as
\begin{equation}
\begin{split}
    &k_{D^0\overline{D}{}^{*0}} = \sqrt{2\mu_{D^0\overline{D}{}^{*0}} E},\\
    &k_{D^+D^{*-}} = \sqrt{2\mu_{D^+D^{*-}} (E - \delta)},\\
    &\kappa_{D^+D^{*-}} = \sqrt{2\mu_{D^+D^{*-}} (\delta - E)},\\
    &\delta = (m_{D^+}+m_{D^{*-}}) - (m_{D^0}+m_{D^{*0}}).\\
\end{split}
\end{equation}
The energy-dependent width $\Gamma(E)$ is defined by
\begin{equation}
    \Gamma(E) = \Gamma_{J/\psi\rho}(E) + \Gamma_{J/\psi\omega}(E) + \Gamma_0,
\end{equation}
where $\Gamma_a$ is the partial width for the channel $a$.
For the $J/\psi\rho$ and $J/\psi\omega$ channels, the dependence on $E$ is defined as follows using the phase space and effective coupling constants, $f_{\rho}$ and $f_{\omega}$~\cite{Kalashnikova2009}:
\begin{equation}
    \Gamma_{J/\psi\rho}(E) = f_{\rho} \int^{M(E)-m_{J/\psi}}_{2m_{\pi}} \frac{dm'}{2\pi} \frac{q(m',E)\Gamma_{\rho}}{(m'-m_{\rho})^2 + \Gamma^2_{\rho}/4},
\end{equation}
\begin{equation}
    \Gamma_{J/\psi\omega}(E) = f_{\omega} \int^{M(E)-m_{J/\psi}}_{3m_{\pi}} \frac{dm'}{2\pi} \frac{q(m',E)\Gamma_{\omega}}{(m'-m_{\omega})^2 + \Gamma^2_{\omega}/4},
\end{equation}
where $\Gamma_{\rho}$ and $\Gamma_{\omega}$ are total widths for the $\rho$ and $\omega$ resonances, respectively.
The upper bound of the integral is set by the difference between 
\begin{equation}
    M(E) = E + (m_{D^0}+m_{D^{*0}})
\end{equation}
and $m_{J/\psi}$.
In each case, $q(m',E)$ is the momentum of the two- or three-pion system in the rest frame of the $X(3872)$: 
\begin{equation}
\begin{split}
    q(m',E) = &\frac{1}{2M(E)}\sqrt{M^2(E)-(m'+m_{J/\psi})^2} \\
    &\times\sqrt{M^2(E)-(m'-m_{J/\psi})^2}.
\end{split}
\end{equation}
The parameter $\Gamma_0$ is the sum of the partial widths of other channels, such as radiative decays. 
In total, this model has five free parameters, $E_f$, $g$, $f_{\rho}$, $f_{\omega}$, and $\Gamma_0$.\par

To obtain stable results in the fit, we apply two constraints, which were also used in the previous study at LHCb~\cite{LHCbFlatte}.
The first is to fix $f_\omega$ so that the branching fraction of the $J/\psi\pi^+\pi^-$ mode and that of the $J/\psi\omega$ mode are equal, consistent with experimental results to date~\cite{jpsiomegaBelle, jpsiomegaBABAR, jpsiomegaBESIII}.
Based on the feature that the area under the lineshape for a channel is proportional to the branching fraction, $f_\omega$ can be uniquely determined by calculating the ratio of the lineshape area in the $J/\psi\pi^+\pi^-$ channel to that in $J/\psi\omega$.
The second is a soft constraint on the ratio of branching fractions between the $J/\psi\pi^+\pi^-$ and $D^0\overline{D}{}^{*0}$ decay modes:
for each of the $B^+$ and $B^0$ modes, the $J/\psi\pi^+\pi^-$ product branching fraction is calculated as follows, and a Gaussian constraint to the measured value~\cite{BRjpsipipi} is included in the fit,
\begin{equation}
\label{fitting:flatte:eq:branchingFractionConstraint}
    \begin{split}
    & {\cal B} (B \to X(3872)K) \times {\cal B}(X(3872) \to J/\psi\pi^+\pi^-) \\
    & = R_{D\overline{D}{}^*} \times {\cal B}(B\to X(3872)K) \times {\cal{B}}(X(3872) \to D^0\overline{D}{}^{*0})\\
    & = \left\{
    \begin{array}{cc}
        (8.61 \pm 0.32) \times 10^{-6} &  \text{for the } B^+ \text{ mode} \\
        (4.1 \pm 1.1) \times 10^{-6} & \text{for the } B^0 \text{ mode} \\
    \end{array}
    \right.,
    \end{split}
\end{equation}
where $R_{D\overline{D}{}^*}$ is the ratio of the lineshape area in the $J/\psi\pi^+\pi^-$ channel to that in $D^0\overline{D}{}^{*0}$, and the $D^0\overline{D}{}^{*0}$ product branching fraction is given by Eq.~\eqref{fitting:flatte:eq:branchingFractionCalc}.\par

There are too few events in our $X(3872)\to D^0\overline{D}{}^{*0}$ sample to simultaneously determine the four remaining parameters.
Therefore, we focus on the parameter regions where scaling behavior was observed at LHCb~\cite{LHCbFlatte}.
We search for the best lineshape fitted to the $M(D^0\overline{D}{}^{*0})$ distribution when the following ratios of parameters are fixed to the values measured at LHCb:
$dg/dE_f$ is fixed to $-15.11~{\rm GeV}^{-1}$, and $f_{\rho}/E_f$ and $\Gamma_0/E_f$ are fixed based on the measurements $f_{\rho}=1.8\times 10^{-3}$ and $\Gamma_0=1.4~{\rm MeV}$, and the assumption $E_f=-7.2~{\rm MeV}$.
Thus, only $g$ is floated as a free parameter.\par

\begin{figure}[bt] 
\begin{center}
\includegraphics[width=\hsize]{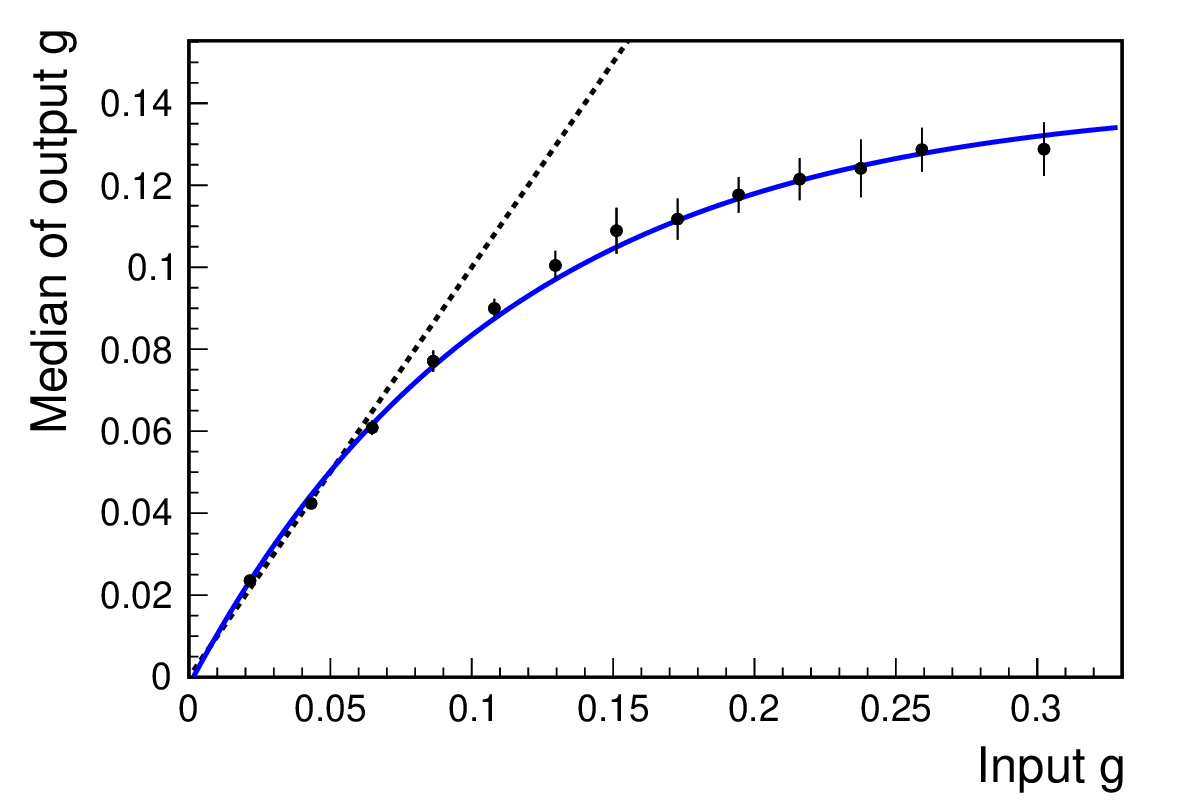}
\caption{\label{result:fig:fitResponseFlatte} 
The median of output values of the coupling constant $g$, as a function of the input $g$, evaluated using pseudo experiments. The dotted black line represents perfect linearity $g_{\rm out} = g_{\rm in}$. The solid blue curve represents the threshold function $g_{\rm out} = 0.14 (1-\exp (-9g_{\rm in}))$.}
\end{center}
\end{figure}

We perform a simultaneous unbinned maximum likelihood fit under the above fit conditions.
The fit results for the data are shown in Fig.~\ref{figID:fitFlatte} and Table~\ref{tbID:fitFlatte}. The fitted $g$ is $0.29^{+2.69}_{-0.15}$, where the uncertainty is statistical. 
Systematic uncertainties are summarized in Table~\ref{tabID:systematicFlatte}. 
The method to evaluate the uncertainties due to the sources (i) to (ix) is the same as in the measurement of the relativistic Breit-Wigner lineshape.
Sources (x) to (xvii) also contribute through the constraint on the branching fraction applied in the fit.
They are evaluated by performing fits after varying each parameter by $\pm 1\sigma$, and adding the resulting changes in quadrature.
Regarding the fitter bias (v), the relationship between input values and medians of output values is evaluated using pseudo experiments, as shown in Fig.~\ref{result:fig:fitResponseFlatte}.
This study shows that for this sample size, $g$ is likely to be underestimated as $g$ increases, with the median of the output values converging to around 0.14.
The main reason is that the lineshape converges to a fixed form for large $g$ (given the assumed ratios for the other parameters), and fits fail, especially in determining an upper statistical uncertainty.

\begin{figure}[bt] 
\begin{center}
\includegraphics[width=\hsize]{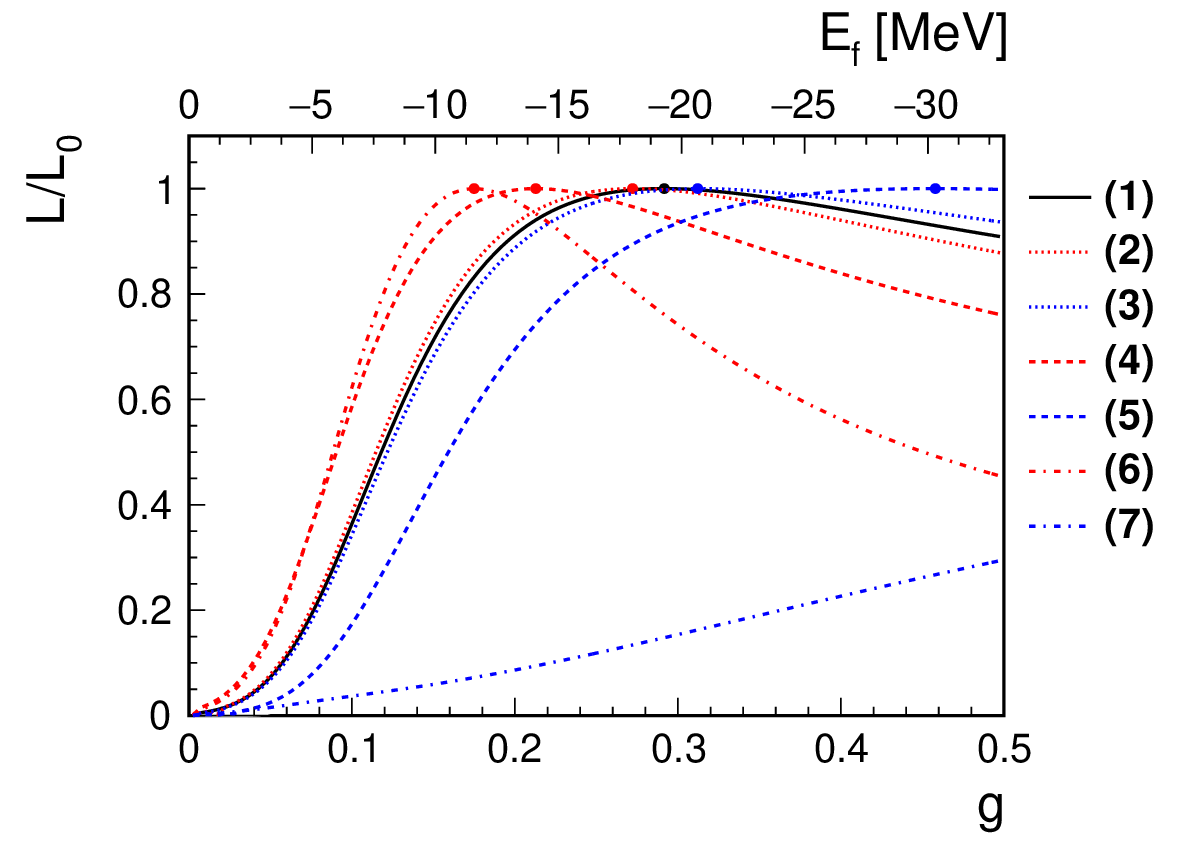}
\caption{\label{result:fig:likelihoodProfile} 
For each of seven parameter sets, the likelihood ratio $L/L_0$ is shown, as a function of the coupling constant $g$,
where $L=L(g)$ is the fitted likelihood and $L_0$ is the likelihood of the best fit for that parameter set. 
The solid black line shows the parameter set (1).
The red and blue dotted lines show parameter sets (2) and (3), respectively.
The red and blue dashed lines show sets (4) and (5), and
the red and blue dot-dashed lines show sets (6) and (7), respectively.
The parameter sets are described in Table~\ref{result:tbID:flatteLowerLimit}. Circles on the lines show the best fit $g$.}
\end{center}
\end{figure}

Since there is no input value for $g$ for which the median of output values is 0.29, we cannot determine a central value for $g$.
We can however set a lower limit.
The likelihood including the systematic uncertainties listed in Table~\ref{tabID:systematicFlatte}, $L(g)$, is shown as the black solid line in Fig.~\ref{result:fig:likelihoodProfile}.
Noting that the curve is asymmetric, with a larger integral above than below the best fit value, we conservatively set the lower limit $g_{\rm lower}$ from
\begin{equation}
\begin{split}
    &\int_{g_{\rm lower}}^{g_{\rm best}}L(g)dg =\\
    & \qquad = \left\{
    \begin{array}{ll}
        \displaystyle 0.8\int_0^{g_{\rm best}} L(g)dg &\text{ for 90\% credibility},\\
        \displaystyle 0.9\int_0^{g_{\rm best}} L(g)dg &\text{ for 95\% credibility},
    \end{array}
    \right.
\end{split}
\end{equation}
where $g_{\rm best}$ denotes the coupling constant at the maximum likelihood.
The effect of fixing $dg/dE_f$, $f_\rho$, and $\Gamma_0$ to the values measured by LHCb is studied by varying each parameter by $\pm1\sigma$. 
Separate curves of the relative likelihood $L/L_0$ for each case are also shown in Fig.~\ref{result:fig:likelihoodProfile}, 
where $L=L(g)$ is the likelihood of the fit and $L_0$ is the likelihood of the best fit for each parameter set. The corresponding fit results and lower limits are summarized in Table~\ref{result:tbID:flatteLowerLimit}.
The $L_0$ values for the different parameter sets vary in a small range around the value for set (1):
the best is favoured by only $1.2\sigma$ relative to set (1), and the worst is disfavoured by $3.4\sigma$.
The loosest lower limit is obtained for the parameter set (6), one of the disfavoured scenarios, where $f_\rho$ is changed by $+1\sigma$. 
We conservatively choose these as the final lower limits for this study:
\[
\begin{split}
g > 0.094 \text{ at 90\% credibility},\\
g > 0.075 \text{ at 95\% credibility}.
\end{split}
\]
These correspond to upper limits of $E_f < -6.2~{\rm MeV}$ at 90\% credibility and $E_f < -5.0~{\rm MeV}$ at 95\% credibility, which are derived from $dg/dE_f = -15.11~{\rm GeV}^{-1}$.\par

We investigate which lineshape model fits the $M(D^{0}\overline{D}{}^{*0})$ distribution better using the test statistic $t=-2\ln ({\cal L}_{\rm BW}/{\cal L}_{\rm Flatte})$.
Here, ${\cal L}_{\rm BW}$ is the best fit likelihood for the Breit-Wigner lineshape, and ${\cal L}_{\rm Flatte}$ is the best likelihood without the $R_{D\overline{D}{}^*}$ constraint term, for the Flatt\'{e} lineshape with parameter set (1).
For data, we obtain $t = -8.5$; \emph{i.e.}\ the Breit-Wigner lineshape is favored.
Based on the $t$ distribution obtained from pseudo experiments, the exclusion level for the Flatt\'{e} lineshape is only $2.2\sigma$;
this level decreases when the systematic uncertainties are taken into account.
Thus, neither lineshape can be excluded.\par

Additionally, the consistency of the two lineshape measurements is confirmed using pseudo experiments.
The Breit-Wigner parameters measured for the data are consistent with those obtained from pseudo experiments generated with the observed \Flatte lineshape.

\begin{table}[tb]
\caption{Results using the \Flatte lineshape: the fitted coupling constant $g$, and the signal yield.}
\label{tbID:fitFlatte}
\begin{tabular}{ccc} \hline \hline
Mode &  $g$ & $N_{\rm sig}$  \\
\hline
Combined & $0.29^{+2.69}_{-0.15}$ & $90.9^{+11.3}_{-15.9}$\\
$X(3872) K^+$ & --- & $77.9^{+\phantom{1}9.6}_{-13.5}$  \\
$X(3872) K^0$ & --- & $13.0^{+\phantom{1}3.0}_{-\phantom{1}2.9}$ \\
\hline \hline
\end{tabular}
\end{table}

\begin{table}[bt]
    \centering
    \caption{Summary of systematic uncertainties for the coupling constant $g$ of the \Flatte lineshape. \label{tabID:systematicFlatte}}
    \setlength{\tabcolsep}{1pt}
    \begin{tabular}{c l  c  }\hline  \hline
        \multicolumn{2}{c}{Source} & $g$  \\ \hline
        (i) & Generic BG PDF & $< O(0.001)$\\
        (ii) & Mass resolution & $-0.011$/$+0.003$\\
        (iii) & Mass dependence of efficiency & $-0.012$/$+0.024$\\
        (iv) & Ratio of broken-signal BG to signal & $-0.007$/$+0.020$\\
        (v) & Fit bias & $-0.000$/$+\infty\phantom{20}\,\,$\\ 
        (vi) & $D^{*0}$ and $D^0$ masses & $\cdots$\\
        (vii) & $D^{*0}$ width & $-0.006$/$+0.001$\\ 
        (viii) & Broken-signal shape for $\overline{D}{}^{*0}\to \overline{D}{}^0\gamma$ & $-0.001$/$+0.002$\\
        (ix) & Signal ratio of $\overline{D}{}^{*0}\to \overline{D}{}^0\gamma$ to $\overline{D}{}^0\pi^0$ & $-0.000$/$+0.004$\\\hline 
        (x)--(xvii) & Branching fraction & $-0.021$/$+0.042$ \\ \hline
        & Total & $-0.029$/$+\infty\phantom{20}\,\,$\\ \hline \hline
    \end{tabular}
\end{table}

\begin{table*}[tb]
\caption{Summary of the seven parameter sets used in the evaluation of lower limits on the coupling constant $g$, showing the $g$ of the best fit, the $g$ lower limits, and corresponding $E_f$ upper limits.
The parameter sets are the center values of $dg/dE_f$, $\Gamma_0$, and $f_\rho$ measured at LHCb~\cite{LHCbFlatte} (1), changing $dg/dE_f$ by $+1\sigma$ (2), changing $dg/dE_f$ by $-1\sigma$ (3), changing $\Gamma_0$ by $+1\sigma$ (4), changing $\Gamma_0$ by $-1\sigma$ (5), changing $f_\rho$ by $+1\sigma$ (6), and changing $f_\rho$ by $-1\sigma$ (7). For the parameter set (7), no lower limit is determined, because no best fit is found in the range $g < 50$.
}
\label{result:tbID:flatteLowerLimit}
    \begin{tabular}{cccccccc} \hline \hline
    Parameter set & (1) & (2)& (3) & (4) & (5) & (6) & (7)\\ \hline
    $dg/dE_f~({\rm GeV}^{-1})$ & $-15.11$ & $-14.95~(+1\sigma)$ & $-15.27~(-1\sigma)$ & $-15.11$ & $-15.11$ & $-15.11$ & $-15.11$ \\
    $\Gamma_0/E_f$ & $-0.19$ & $-0.19$ & $-0.19$ & $-0.29~(+1\sigma)$ & $-0.09~(-1\sigma)$ & $-0.19$ & $-0.19$\\
    $f_\rho/E_f~({\rm GeV}^{-1})$ & $-0.25$ & $-0.25$ & $-0.25$ & $-0.25$ & $-0.25$ & $-0.38~(+1\sigma)$ & $-0.12~(-1\sigma)$\\ \hline
    $g$ of best fit & $0.29$ & $0.27$ & $0.31$ & $0.21$ & $0.46$ & $0.17$ & $> 50$ \\ \hline
    $g$ lower limit at 90\% C.L. & $>0.143$ & $>0.136$ & $>0.151$ & $>0.105$ & $> 0.212$ & $> 0.094$ & --- \\ 
    \hspace{5.7em} at 95\% C.L. & $>0.113$ & $>0.108$ & $>0.119$ & $>0.082$ & $> 0.167$ & $> 0.075$ & --- \\ \hline
    $E_f$ upper limit at 90\% C.L.~(${\rm MeV}$) & $<-9.5$ & $< -9.0$ & $< -10.0$ & $<-6.9$ & $<-14.0$ & $<-6.2$ & ---\\ 
    \hspace{6.6em} at 95\% C.L.~(${\rm MeV}$) & $<-7.6$ & $< -7.2$ & $< -7.9$ & $<-5.5$ & $<-11.1$ & $<-5.0$ & ---\\ \hline \hline
\end{tabular}
\end{table*}

\section{Discussion and Conclusion}
\label{secID:conclusion}

In this paper, we examine the $X(3872) \to D^0\overline{D}{}^{*0}$ lineshape using the full Belle dataset.
When fitting with a relativistic Breit-Wigner lineshape, the mass and width parameters are measured to be
\begin{equation*}
\begin{split}
m_{\rm BW} &= 3873.71 ^{+0.56}_{-0.50} ({\rm stat}) \pm 0.13 ({\rm syst}) ~{\rm MeV}/c^2,\\
\Gamma_{\rm BW} &= 5.2 ^{+2.2}_{-1.5} ({\rm stat}) \pm 0.4 ({\rm syst})~{\rm MeV}.
\end{split}
\end{equation*}
The difference between $m_{\rm BW}$ and the $D^0\overline{D}{}^{*0}$ threshold is found to be $2.02^{+0.56}_{-0.50} ({\rm stat}) \pm 0.08 ({\rm syst})~{\rm MeV}/c^2$.
These values are in good agreement with those measured in previous studies of the $D^0\overline{D}{}^{*0}$ decay~{\cite{BaBarDDstar, Aushev2010}}, 
and the precision of the measurement is improved by at least 22\%.
The measured branching fractions are as follows:
\begin{equation*}
\begin{split}
    {\cal B}& (B^+\to X(3872)K^+) \times {\cal{B}}(X(3872) \to D^0\overline{D}{}^{*0}) \\
    &  \qquad = (0.97 ^{+0.21}_{-0.18} ({\rm stat}) \pm 0.10 ({\rm syst})) \times 10^{-4},\\
    {\cal B}&(B^0 \to X(3872)K^0) \times {\cal{B}}(X(3872) \to D^0\overline{D}{}^{*0}) \\
    & \qquad = (1.30 ^{+0.36}_{-0.31} ({\rm stat}) ^{+0.12}_{-0.07} ({\rm syst})) \times 10^{-4}.\\
\end{split}
\end{equation*}
This is the first measurement of $X(3872)$ production in $B^0$ decays with more than 5$\sigma$ significance.
The ratio of the branching fractions is determined to be 
\begin{equation*}
    \frac{{\cal B}(B^0\to X(3872)K^0)}{{\cal B}(B^+ \to X(3872)K^+)} = 1.34^{+0.47}_{-0.40} ({\rm stat}) ^{+0.10}_{-0.12} ({\rm syst}).
\end{equation*}
These results are in good agreement with those of previous studies~\cite{Gokhroo2006, BaBarDDstar, Aushev2010}.\par

We compare these results with the analysis of the Breit-Wigner lineshape using the $J/\psi\pi^+\pi^-$ decay mode.
The measured Breit-Wigner mass is significantly higher than the $D^0\overline{D}{}^{*0}$ threshold, while the world-average mass with the $J/\psi\pi^+\pi^-$ decay is consistent with the threshold.
The measured width and ratio ${\cal B}(B^0\to X(3872)K^0)/{\cal B}(B^+ \to X(3872)K^+)$ are shifted from the average with the $J/\psi\pi^+\pi^-$ decay by $2.6\sigma$ and $2.0\sigma$, respectively~\cite{RXJpsipipi}.
In previous studies of the $D^0\overline{D}{}^{*0}$ decay, it has also been seen that these properties in the $X(3872) \to D^0\overline{D}{}^{*0}$ decay mode differ from those in $J/\psi\pi^+\pi^-$.
\par

We also fit the lineshape using a Flatt\'{e}-inspired parameterization.
With sufficient data, such a model could be used to simultaneously describe the lineshapes of the decays to the $J/\psi\pi^+\pi^-$ and $D^0\bar{D}^{*0}$ final states.
Given the limited size of the $D^0\overline{D}{}^{*0}$ data sample at Belle, and the scaling behaviour observed in the LHCb study of $J/\psi\pi^+\pi^-$,
we set various ratios of parameters to their LHCb values, and fit with the coupling constant to the $D\overline{D}{}^*$ channel, $g$, as the undetermined parameter.
We find that the fitted value of $g$ is in a region that is relatively insensitive to the underlying value.
We determine its lower limits to be
\[
\begin{split}
g > 0.094 \text{ at 90\% credibility},\\
g > 0.075 \text{ at 95\% credibility}.
\end{split}
\]
These correspond to upper limits of $E_f < -6.2~{\rm MeV}$ at 90\% credibility and $E_f < -5.0~{\rm MeV}$ at 95\% credibility,
which are slightly more stringent than the LHCb measurement, $-270~{\rm MeV} < E_f <-2.0~{\rm MeV}$~\cite{LHCbFlatte}.
This suggests that analysis using $D^0\overline{D}{}^{*0}$ can indeed complement the study of the $J/\psi\pi^+\pi^-$ mode in this framework.
The limit includes the solution $E_f=-7.2~{\rm MeV}$ assumed in the scattering amplitude analysis at LHCb.
There is still uncertainty in the pole positions of the scattering amplitude, because the limit is not especially stringent.
\par

Both Breit-Wigner and \Flatte lineshapes fit the invariant mass distribution obtained from the data.
Finally, we examine which lineshape model best fits the invariant mass distribution.
Based on the likelihood ratio from the fits, the Breit-Wigner lineshape is favored, but the \Flatte lineshape is not excluded.\par

Analysis of the large dataset expected from Belle II will be important, because the statistical uncertainty dominates in both of the lineshape measurements.
In the \Flatte study, it is essential to reduce systematic uncertainty due to the fit bias and the measurement of parameters $f_\rho$ and $\Gamma_0$.
Increasing the size of the data sample will also reduce these uncertainties.
In addition, a simultaneous fit of the $J/\psi\pi^+\pi^-$ and $D^0\overline{D}{}^{*0}$ decay modes will also be useful, because the ratio of branching fractions can further constrain the parameters.
In such a fit, the most adequate samples would be an exclusive $B\to X(3872)(\to J/\psi\pi^+\pi^-)K$ sample at LHCb and a $B\to X(3872)(\to D^0\overline{D}{}^{*0})K$ sample at Belle II.
Such an analysis could fully determine the lineshape in the coupled-channel framework, and greatly contribute to determining the internal structure.


\section*{acknowledgements} 

This work, based on data collected using the Belle detector, which was
operated until June 2010, was supported by 
the Ministry of Education, Culture, Sports, Science, and
Technology (MEXT) of Japan, the Japan Society for the 
Promotion of Science (JSPS) including in 
particular the Grant-in-Aid for JSPS Fellows (No.19J23314), 
the Tau-Lepton Physics Research Center of Nagoya University,
and the Advanced Science Research Center of Japan Atomic Energy Agency; 
the Australian Research Council including grants
DP180102629, 
DP170102389, 
DP170102204, 
DE220100462, 
DP150103061, 
FT130100303; 
Austrian Federal Ministry of Education, Science and Research (FWF) and
FWF Austrian Science Fund No.~P~31361-N36;
the National Natural Science Foundation of China under Contracts
No.~11675166,  
No.~11705209;  
No.~11975076;  
No.~12135005;  
No.~12175041;  
No.~12161141008; 
Key Research Program of Frontier Sciences, Chinese Academy of Sciences (CAS), Grant No.~QYZDJ-SSW-SLH011; 
Project ZR2022JQ02 supported by Shandong Provincial Natural Science Foundation;
the Ministry of Education, Youth and Sports of the Czech
Republic under Contract No.~LTT17020;
the Czech Science Foundation Grant No. 22-18469S;
Horizon 2020 ERC Advanced Grant No.~884719 and ERC Starting Grant No.~947006 ``InterLeptons'' (European Union);
the Carl Zeiss Foundation, the Deutsche Forschungsgemeinschaft, the
Excellence Cluster Universe, and the VolkswagenStiftung;
the Department of Atomic Energy (Project Identification No. RTI 4002) and the Department of Science and Technology of India; 
the Istituto Nazionale di Fisica Nucleare of Italy; 
National Research Foundation (NRF) of Korea Grant
Nos.~2016R1\-D1A1B\-02012900, 2018R1\-A2B\-3003643,
2018R1\-A6A1A\-06024970, RS\-2022\-00197659,
2019R1\-I1A3A\-01058933, 2021R1\-A6A1A\-03043957,
2021R1\-F1A\-1060423, 2021R1\-F1A\-1064008, 2022R1\-A2C\-1003993;
Radiation Science Research Institute, Foreign Large-size Research Facility Application Supporting project, the Global Science Experimental Data Hub Center of the Korea Institute of Science and Technology Information and KREONET/GLORIAD;
the Polish Ministry of Science and Higher Education and 
the National Science Center;
the Ministry of Science and Higher Education of the Russian Federation, Agreement 14.W03.31.0026, 
and the HSE University Basic Research Program, Moscow; 
University of Tabuk research grants
S-1440-0321, S-0256-1438, and S-0280-1439 (Saudi Arabia);
the Slovenian Research Agency Grant Nos. J1-9124 and P1-0135;
Ikerbasque, Basque Foundation for Science, Spain;
the Swiss National Science Foundation; 
the Ministry of Education and the Ministry of Science and Technology of Taiwan;
and the United States Department of Energy and the National Science Foundation.
These acknowledgements are not to be interpreted as an endorsement of any
statement made by any of our institutes, funding agencies, governments, or
their representatives.
We thank the KEKB group for the excellent operation of the
accelerator; the KEK cryogenics group for the efficient
operation of the solenoid; and the KEK computer group and the Pacific Northwest National
Laboratory (PNNL) Environmental Molecular Sciences Laboratory (EMSL)
computing group for strong computing support; and the National
Institute of Informatics, and Science Information NETwork 6 (SINET6) for
valuable network support.

\bibliographystyle{apsrev4-2} 
\bibliography{main} 

\begin{thebibliography}{50}%
\makeatletter
\providecommand \@ifxundefined [1]{%
 \@ifx{#1\undefined}
}%
\providecommand \@ifnum [1]{%
 \ifnum #1\expandafter \@firstoftwo
 \else \expandafter \@secondoftwo
 \fi
}%
\providecommand \@ifx [1]{%
 \ifx #1\expandafter \@firstoftwo
 \else \expandafter \@secondoftwo
 \fi
}%
\providecommand \natexlab [1]{#1}%
\providecommand \enquote  [1]{``#1''}%
\providecommand \bibnamefont  [1]{#1}%
\providecommand \bibfnamefont [1]{#1}%
\providecommand \citenamefont [1]{#1}%
\providecommand \href@noop [0]{\@secondoftwo}%
\providecommand \href [0]{\begingroup \@sanitize@url \@href}%
\providecommand \@href[1]{\@@startlink{#1}\@@href}%
\providecommand \@@href[1]{\endgroup#1\@@endlink}%
\providecommand \@sanitize@url [0]{\catcode `\\12\catcode `\$12\catcode
  `\&12\catcode `\#12\catcode `\^12\catcode `\_12\catcode `\%12\relax}%
\providecommand \@@startlink[1]{}%
\providecommand \@@endlink[0]{}%
\providecommand \url  [0]{\begingroup\@sanitize@url \@url }%
\providecommand \@url [1]{\endgroup\@href {#1}{\urlprefix }}%
\providecommand \urlprefix  [0]{URL }%
\providecommand \Eprint [0]{\href }%
\providecommand \doibase [0]{https://doi.org/}%
\providecommand \selectlanguage [0]{\@gobble}%
\providecommand \bibinfo  [0]{\@secondoftwo}%
\providecommand \bibfield  [0]{\@secondoftwo}%
\providecommand \translation [1]{[#1]}%
\providecommand \BibitemOpen [0]{}%
\providecommand \bibitemStop [0]{}%
\providecommand \bibitemNoStop [0]{.\EOS\space}%
\providecommand \EOS [0]{\spacefactor3000\relax}%
\providecommand \BibitemShut  [1]{\csname bibitem#1\endcsname}%
\let\auto@bib@innerbib\@empty
\bibitem [{\citenamefont {Workman}\ \emph {et~al.}(2022)\citenamefont {Workman}
  \emph {et~al.}}]{PDG2020}%
  \BibitemOpen
  \bibfield  {author} {\bibinfo {author} {\bibfnamefont {R.~L.}\ \bibnamefont
  {Workman}} \emph {et~al.} (\bibinfo {collaboration} {Particle Data Group}),\
  }\href {https://doi.org/10.1093/ptep/ptac097} {\bibfield  {journal} {\bibinfo
   {journal} {Prog. Theor. Exp. Phys.}\ }\textbf {\bibinfo {volume} {2022}},\
  \bibinfo {pages} {083C01} (\bibinfo {year} {2022})}\BibitemShut {NoStop}%
\bibitem [{\citenamefont {Choi}\ \emph {et~al.}(2003)\citenamefont {Choi} \emph
  {et~al.}}]{XDiscoveryBelle}%
  \BibitemOpen
  \bibfield  {author} {\bibinfo {author} {\bibfnamefont {S.-K.}\ \bibnamefont
  {Choi}} \emph {et~al.} (\bibinfo {collaboration} {Belle Collaboration}),\
  }\href {https://doi.org/10.1103/PhysRevLett.91.262001} {\bibfield  {journal}
  {\bibinfo  {journal} {Phys. Rev. Lett.}\ }\textbf {\bibinfo {volume} {91}},\
  \bibinfo {pages} {262001} (\bibinfo {year} {2003})}\BibitemShut {NoStop}%
\bibitem [{\citenamefont {Abazov}\ \emph {et~al.}(2004)\citenamefont {Abazov}
  \emph {et~al.}}]{XDiscoveryD0}%
  \BibitemOpen
  \bibfield  {author} {\bibinfo {author} {\bibfnamefont {V.~M.}\ \bibnamefont
  {Abazov}} \emph {et~al.} (\bibinfo {collaboration} {D0 Collaboration}),\
  }\href {https://doi.org/10.1103/PhysRevLett.93.162002} {\bibfield  {journal}
  {\bibinfo  {journal} {Phys. Rev. Lett.}\ }\textbf {\bibinfo {volume} {93}},\
  \bibinfo {pages} {162002} (\bibinfo {year} {2004})}\BibitemShut {NoStop}%
\bibitem [{\citenamefont {Aubert}\ \emph {et~al.}(2005)\citenamefont {Aubert}
  \emph {et~al.}}]{XDiscoveryBABAR}%
  \BibitemOpen
  \bibfield  {author} {\bibinfo {author} {\bibfnamefont {B.}~\bibnamefont
  {Aubert}} \emph {et~al.} (\bibinfo {collaboration} {\BABAR Collaboration}),\
  }\href {https://doi.org/10.1103/PhysRevD.71.071103} {\bibfield  {journal}
  {\bibinfo  {journal} {Phys. Rev. D}\ }\textbf {\bibinfo {volume} {71}},\
  \bibinfo {pages} {071103} (\bibinfo {year} {2005})}\BibitemShut {NoStop}%
\bibitem [{\citenamefont {Acosta}\ \emph {et~al.}(2004)\citenamefont {Acosta}
  \emph {et~al.}}]{XDiscoveryCDF}%
  \BibitemOpen
  \bibfield  {author} {\bibinfo {author} {\bibfnamefont {D.}~\bibnamefont
  {Acosta}} \emph {et~al.} (\bibinfo {collaboration} {CDF II Collaboration}),\
  }\href {https://doi.org/10.1103/PhysRevLett.93.072001} {\bibfield  {journal}
  {\bibinfo  {journal} {Phys. Rev. Lett.}\ }\textbf {\bibinfo {volume} {93}},\
  \bibinfo {pages} {072001} (\bibinfo {year} {2004})}\BibitemShut {NoStop}%
\bibitem [{\citenamefont {Aaij}\ \emph {et~al.}(2012)\citenamefont {Aaij} \emph
  {et~al.}}]{XDiscoveryLHCb}%
  \BibitemOpen
  \bibfield  {author} {\bibinfo {author} {\bibfnamefont {R.}~\bibnamefont
  {Aaij}} \emph {et~al.} (\bibinfo {collaboration} {LHCb Collaboration}),\
  }\href {https://doi.org/10.1140/epjc/s10052-012-1972-7} {\bibfield  {journal}
  {\bibinfo  {journal} {Eur. Phys. J. C}\ }\textbf {\bibinfo {volume} {72}}
  (\bibinfo {year} {2012})}\BibitemShut {NoStop}%
\bibitem [{\citenamefont {Ablikim}\ \emph {et~al.}(2014)\citenamefont {Ablikim}
  \emph {et~al.}}]{XDiscoveryBESIII}%
  \BibitemOpen
  \bibfield  {author} {\bibinfo {author} {\bibfnamefont {M.}~\bibnamefont
  {Ablikim}} \emph {et~al.} (\bibinfo {collaboration} {BESIII Collaboration}),\
  }\href {https://doi.org/10.1103/PhysRevLett.112.092001} {\bibfield  {journal}
  {\bibinfo  {journal} {Phys. Rev. Lett.}\ }\textbf {\bibinfo {volume} {112}},\
  \bibinfo {pages} {092001} (\bibinfo {year} {2014})}\BibitemShut {NoStop}%
\bibitem [{Jps()}]{JpsiomegaDiscovery}%
  \BibitemOpen
  \href@noop {} {}\bibinfo {note} {{K. Abe {\it et al.} (Belle Collaboration),
  \href{https://arxiv.org/abs/hep-ex/0505037}{arXiv:hep-ex/0505037};
  P.~del~Amo~Sanchez {\it et al.} (\BABAR Collaboration),
  \href{https://link.aps.org/doi/10.1103/PhysRevD.82.011101}{Phys. Rev. D {\bf
  82}, 011101 (2010)}; M. Ablikim {\it et al.} (BESIII Collaboration) \href
  {https://link.aps.org/doi/10.1103/PhysRevLett.122.232002}{Phys. Rev. Lett.
  {\bf 122}, 232002 (2019)}}}\BibitemShut {NoStop}%
\bibitem [{rad()}]{radiativeDecaysDiscovery}%
  \BibitemOpen
  \href@noop {} {}\bibinfo {note} {{ B. Aubert {\it et al.} (\BABAR
  Collaboration),
  \href{https://link.aps.org/doi/10.1103/PhysRevLett.102.132001}{Phys. Rev.
  Lett. {\bf 102}, 132001 (2009)}; V. Bhardwaj {\it et al.} (Belle
  Collaboration),
  \href{https://link.aps.org/doi/10.1103/PhysRevLett.107.091803}{Phys. Rev.
  Lett. {\bf 107}, 091803 (2011)}; R. Aaij {\it et al.} (LHCb Collaboration),
  \href{https://www.sciencedirect.com/science/article/pii/S0550321314001941}{Nuc.
  Phys. B{\bf 886}, 665 (2014)}; M. Ablikim {\it et al.} (BESIII
  Collaboration),
  \href{https://link.aps.org/doi/10.1103/PhysRevLett.124.242001}{Phys. Rev.
  Lett. {\bf 124}, 242001 (2020)} }}\BibitemShut {NoStop}%
\bibitem [{\citenamefont {Aubert}\ \emph
  {et~al.}(2008{\natexlab{a}})\citenamefont {Aubert} \emph
  {et~al.}}]{BaBarDDstar}%
  \BibitemOpen
  \bibfield  {author} {\bibinfo {author} {\bibfnamefont {B.}~\bibnamefont
  {Aubert}} \emph {et~al.} (\bibinfo {collaboration} {\BABAR Collaboration}),\
  }\href {https://doi.org/10.1103/PhysRevD.77.011102} {\bibfield  {journal}
  {\bibinfo  {journal} {Phys. Rev. D}\ }\textbf {\bibinfo {volume} {77}},\
  \bibinfo {pages} {011102} (\bibinfo {year} {2008}{\natexlab{a}})}\BibitemShut
  {NoStop}%
\bibitem [{\citenamefont {Aushev}\ \emph {et~al.}(2010)\citenamefont {Aushev}
  \emph {et~al.}}]{Aushev2010}%
  \BibitemOpen
  \bibfield  {author} {\bibinfo {author} {\bibfnamefont {T.}~\bibnamefont
  {Aushev}} \emph {et~al.} (\bibinfo {collaboration} {Belle Collaboration}),\
  }\href {https://doi.org/10.1103/PhysRevD.81.031103} {\bibfield  {journal}
  {\bibinfo  {journal} {Phys. Rev. D}\ }\textbf {\bibinfo {volume} {81}},\
  \bibinfo {pages} {031103} (\bibinfo {year} {2010})}\BibitemShut {NoStop}%
\bibitem [{\citenamefont {Gokhroo}\ \emph {et~al.}(2006)\citenamefont {Gokhroo}
  \emph {et~al.}}]{Gokhroo2006}%
  \BibitemOpen
  \bibfield  {author} {\bibinfo {author} {\bibfnamefont {G.}~\bibnamefont
  {Gokhroo}} \emph {et~al.} (\bibinfo {collaboration} {Belle Collaboration}),\
  }\href {https://doi.org/10.1103/PhysRevLett.97.162002} {\bibfield  {journal}
  {\bibinfo  {journal} {Phys. Rev. Lett.}\ }\textbf {\bibinfo {volume} {97}},\
  \bibinfo {pages} {162002} (\bibinfo {year} {2006})}\BibitemShut {NoStop}%
\bibitem [{\citenamefont {Bhardwaj}\ \emph {et~al.}(2019)\citenamefont
  {Bhardwaj} \emph {et~al.}}]{chicJpi0Discovery}%
  \BibitemOpen
  \bibfield  {author} {\bibinfo {author} {\bibfnamefont {V.}~\bibnamefont
  {Bhardwaj}} \emph {et~al.} (\bibinfo {collaboration} {Belle Collaboration}),\
  }\href {https://doi.org/10.1103/PhysRevD.99.111101} {\bibfield  {journal}
  {\bibinfo  {journal} {Phys. Rev. D}\ }\textbf {\bibinfo {volume} {99}},\
  \bibinfo {pages} {111101} (\bibinfo {year} {2019})}\BibitemShut {NoStop}%
\bibitem [{\citenamefont {Aaij}\ \emph {et~al.}(2013)\citenamefont {Aaij} \emph
  {et~al.}}]{LHCbXJpc1}%
  \BibitemOpen
  \bibfield  {author} {\bibinfo {author} {\bibfnamefont {R.}~\bibnamefont
  {Aaij}} \emph {et~al.} (\bibinfo {collaboration} {LHCb Collaboration}),\
  }\href {https://doi.org/10.1103/PhysRevLett.110.222001} {\bibfield  {journal}
  {\bibinfo  {journal} {Phys. Rev. Lett.}\ }\textbf {\bibinfo {volume} {110}},\
  \bibinfo {pages} {222001} (\bibinfo {year} {2013})}\BibitemShut {NoStop}%
\bibitem [{\citenamefont {Aaij}\ \emph {et~al.}(2015)\citenamefont {Aaij} \emph
  {et~al.}}]{LHCbXJpc2}%
  \BibitemOpen
  \bibfield  {author} {\bibinfo {author} {\bibfnamefont {R.}~\bibnamefont
  {Aaij}} \emph {et~al.} (\bibinfo {collaboration} {LHCb Collaboration}),\
  }\href {https://doi.org/10.1103/PhysRevD.92.011102} {\bibfield  {journal}
  {\bibinfo  {journal} {Phys. Rev. D}\ }\textbf {\bibinfo {volume} {92}},\
  \bibinfo {pages} {011102} (\bibinfo {year} {2015})}\BibitemShut {NoStop}%
\bibitem [{\citenamefont {Törnqvist}(2004)}]{molecular1}%
  \BibitemOpen
  \bibfield  {author} {\bibinfo {author} {\bibfnamefont {N.~A.}\ \bibnamefont
  {Törnqvist}},\ }\href
  {https://doi.org/https://doi.org/10.1016/j.physletb.2004.03.077} {\bibfield
  {journal} {\bibinfo  {journal} {Phys. Lett. B}\ }\textbf {\bibinfo {volume}
  {590}},\ \bibinfo {pages} {209} (\bibinfo {year} {2004})}\BibitemShut
  {NoStop}%
\bibitem [{\citenamefont {Swanson}(2004)}]{molecular2}%
  \BibitemOpen
  \bibfield  {author} {\bibinfo {author} {\bibfnamefont {E.~S.}\ \bibnamefont
  {Swanson}},\ }\href
  {https://doi.org/https://doi.org/10.1016/j.physletb.2004.03.033} {\bibfield
  {journal} {\bibinfo  {journal} {Phys. Lett. B}\ }\textbf {\bibinfo {volume}
  {588}},\ \bibinfo {pages} {189} (\bibinfo {year} {2004})}\BibitemShut
  {NoStop}%
\bibitem [{\citenamefont {Wong}(2004)}]{molecular3}%
  \BibitemOpen
  \bibfield  {author} {\bibinfo {author} {\bibfnamefont {C.-Y.}\ \bibnamefont
  {Wong}},\ }\href {https://doi.org/10.1103/PhysRevC.69.055202} {\bibfield
  {journal} {\bibinfo  {journal} {Phys. Rev. C}\ }\textbf {\bibinfo {volume}
  {69}},\ \bibinfo {pages} {055202} (\bibinfo {year} {2004})}\BibitemShut
  {NoStop}%
\bibitem [{\citenamefont {Gamermann}\ and\ \citenamefont
  {Oset}(2009)}]{molecular4}%
  \BibitemOpen
  \bibfield  {author} {\bibinfo {author} {\bibfnamefont {D.}~\bibnamefont
  {Gamermann}}\ and\ \bibinfo {author} {\bibfnamefont {E.}~\bibnamefont
  {Oset}},\ }\href {https://doi.org/10.1103/PhysRevD.80.014003} {\bibfield
  {journal} {\bibinfo  {journal} {Phys. Rev. D}\ }\textbf {\bibinfo {volume}
  {80}},\ \bibinfo {pages} {014003} (\bibinfo {year} {2009})}\BibitemShut
  {NoStop}%
\bibitem [{\citenamefont {Takizawa}\ and\ \citenamefont
  {Takeuchi}(2013)}]{admixtureCCandMolecule}%
  \BibitemOpen
  \bibfield  {author} {\bibinfo {author} {\bibfnamefont {M.}~\bibnamefont
  {Takizawa}}\ and\ \bibinfo {author} {\bibfnamefont {S.}~\bibnamefont
  {Takeuchi}},\ }\href {https://doi.org/10.1093/ptep/ptt063} {\bibfield
  {journal} {\bibinfo  {journal} {Prog. Theor. Exp. Phys.}\ }\textbf {\bibinfo
  {volume} {2013}},\ \bibinfo {pages} {093D01} (\bibinfo {year}
  {2013})}\BibitemShut {NoStop}%
\bibitem [{\citenamefont {Maiani}\ \emph {et~al.}(2005)\citenamefont {Maiani},
  \citenamefont {Piccinini}, \citenamefont {Polosa},\ and\ \citenamefont
  {Riquer}}]{tetraquark}%
  \BibitemOpen
  \bibfield  {author} {\bibinfo {author} {\bibfnamefont {L.}~\bibnamefont
  {Maiani}}, \bibinfo {author} {\bibfnamefont {F.}~\bibnamefont {Piccinini}},
  \bibinfo {author} {\bibfnamefont {A.~D.}\ \bibnamefont {Polosa}},\ and\
  \bibinfo {author} {\bibfnamefont {V.}~\bibnamefont {Riquer}},\ }\href
  {https://doi.org/10.1103/PhysRevD.71.014028} {\bibfield  {journal} {\bibinfo
  {journal} {Phys. Rev. D}\ }\textbf {\bibinfo {volume} {71}},\ \bibinfo
  {pages} {014028} (\bibinfo {year} {2005})}\BibitemShut {NoStop}%
\bibitem [{\citenamefont {Hanhart}\ \emph {et~al.}(2007)\citenamefont
  {Hanhart}, \citenamefont {Kalashnikova}, \citenamefont {Kudryavtsev},\ and\
  \citenamefont {Nefediev}}]{Hanhart2007}%
  \BibitemOpen
  \bibfield  {author} {\bibinfo {author} {\bibfnamefont {C.}~\bibnamefont
  {Hanhart}}, \bibinfo {author} {\bibfnamefont {Y.~S.}\ \bibnamefont
  {Kalashnikova}}, \bibinfo {author} {\bibfnamefont {A.~E.}\ \bibnamefont
  {Kudryavtsev}},\ and\ \bibinfo {author} {\bibfnamefont {A.~V.}\ \bibnamefont
  {Nefediev}},\ }\href {https://doi.org/10.1103/PhysRevD.76.034007} {\bibfield
  {journal} {\bibinfo  {journal} {Phys. Rev. D}\ }\textbf {\bibinfo {volume}
  {76}},\ \bibinfo {pages} {034007} (\bibinfo {year} {2007})}\BibitemShut
  {NoStop}%
\bibitem [{\citenamefont {Braaten}\ and\ \citenamefont
  {Lu}(2007)}]{Braaten2007}%
  \BibitemOpen
  \bibfield  {author} {\bibinfo {author} {\bibfnamefont {E.}~\bibnamefont
  {Braaten}}\ and\ \bibinfo {author} {\bibfnamefont {M.}~\bibnamefont {Lu}},\
  }\href {https://doi.org/10.1103/PhysRevD.76.094028} {\bibfield  {journal}
  {\bibinfo  {journal} {Phys. Rev. D}\ }\textbf {\bibinfo {volume} {76}},\
  \bibinfo {pages} {094028} (\bibinfo {year} {2007})}\BibitemShut {NoStop}%
\bibitem [{\citenamefont {Bugg}(2008)}]{Bugg2008}%
  \BibitemOpen
  \bibfield  {author} {\bibinfo {author} {\bibfnamefont {D.~V.}\ \bibnamefont
  {Bugg}},\ }\href {https://doi.org/10.1088/0954-3899/35/7/075005} {\bibfield
  {journal} {\bibinfo  {journal} {J. Phys. G}\ }\textbf {\bibinfo {volume}
  {35}},\ \bibinfo {pages} {075005} (\bibinfo {year} {2008})}\BibitemShut
  {NoStop}%
\bibitem [{\citenamefont {Aaij}\ \emph {et~al.}()\citenamefont {Aaij} \emph
  {et~al.}}]{LHCbBRJpsipipi2020}%
  \BibitemOpen
  \bibfield  {author} {\bibinfo {author} {\bibfnamefont {R.}~\bibnamefont
  {Aaij}} \emph {et~al.} (\bibinfo {collaboration} {LHCb Collaboration}),\
  }\href {https://doi.org/10.1007/jhep08(2020)123} {\bibfield  {journal}
  {\bibinfo  {journal} {J. High Energy Phys.}\ }\textbf {\bibinfo {volume}
  {2020}},\ \bibinfo {pages} {123~(2020)}}\BibitemShut {NoStop}%
\bibitem [{\citenamefont {Aaij}\ \emph {et~al.}(2020)\citenamefont {Aaij} \emph
  {et~al.}}]{LHCbFlatte}%
  \BibitemOpen
  \bibfield  {author} {\bibinfo {author} {\bibfnamefont {R.}~\bibnamefont
  {Aaij}} \emph {et~al.} (\bibinfo {collaboration} {LHCb Collaboration}),\
  }\href {https://doi.org/10.1103/PhysRevD.102.092005} {\bibfield  {journal}
  {\bibinfo  {journal} {Phys. Rev. D}\ }\textbf {\bibinfo {volume} {102}},\
  \bibinfo {pages} {092005} (\bibinfo {year} {2020})}\BibitemShut {NoStop}%
\bibitem [{\citenamefont {Kalashnikova}\ and\ \citenamefont
  {Nefediev}(2009)}]{Kalashnikova2009}%
  \BibitemOpen
  \bibfield  {author} {\bibinfo {author} {\bibfnamefont {Y.~S.}\ \bibnamefont
  {Kalashnikova}}\ and\ \bibinfo {author} {\bibfnamefont {A.~V.}\ \bibnamefont
  {Nefediev}},\ }\href {https://doi.org/10.1103/PhysRevD.80.074004} {\bibfield
  {journal} {\bibinfo  {journal} {Phys. Rev. D}\ }\textbf {\bibinfo {volume}
  {80}},\ \bibinfo {pages} {074004} (\bibinfo {year} {2009})}\BibitemShut
  {NoStop}%
\bibitem [{\citenamefont {Baru}\ \emph {et~al.}(2005)\citenamefont {Baru} \emph
  {et~al.}}]{Baru2005}%
  \BibitemOpen
  \bibfield  {author} {\bibinfo {author} {\bibfnamefont {V.}~\bibnamefont
  {Baru}} \emph {et~al.},\ }\href {https://doi.org/10.1140/epja/i2004-10105-x}
  {\bibfield  {journal} {\bibinfo  {journal} {Eur. Phys. J. A}\ }\textbf
  {\bibinfo {volume} {23}},\ \bibinfo {pages} {523} (\bibinfo {year}
  {2005})}\BibitemShut {NoStop}%
\bibitem [{\citenamefont {Kurokawa}\ and\ \citenamefont
  {Kikutani}(2003)}]{KEKB}%
  \BibitemOpen
  \bibfield  {author} {\bibinfo {author} {\bibfnamefont {S.}~\bibnamefont
  {Kurokawa}}\ and\ \bibinfo {author} {\bibfnamefont {E.}~\bibnamefont
  {Kikutani}},\ }\href
  {https://doi.org/https://doi.org/10.1016/S0168-9002(02)01771-0} {\bibfield
  {journal} {\bibinfo  {journal} {Nucl. Instrum. Methods Phys. Res., Sect A}\
  }\textbf {\bibinfo {volume} {499}},\ \bibinfo {pages} {1} (\bibinfo {year}
  {2003})},\ \bibinfo {note} {and other papers included in this
  volume.}\BibitemShut {Stop}%
\bibitem [{\citenamefont {Abe}\ \emph {et~al.}(2013)\citenamefont {Abe} \emph
  {et~al.}}]{achivementKEKB}%
  \BibitemOpen
  \bibfield  {author} {\bibinfo {author} {\bibfnamefont {T.}~\bibnamefont
  {Abe}} \emph {et~al.},\ }\href {https://doi.org/10.1093/ptep/pts102}
  {\bibfield  {journal} {\bibinfo  {journal} {Prog. Theor. Exp. Phys.}\
  }\textbf {\bibinfo {volume} {2013}},\ \bibinfo {pages} {03A001} (\bibinfo
  {year} {2013})}\BibitemShut {NoStop}%
\bibitem [{\citenamefont {Abashian}\ \emph {et~al.}(2002)\citenamefont
  {Abashian} \emph {et~al.}}]{Belle}%
  \BibitemOpen
  \bibfield  {author} {\bibinfo {author} {\bibfnamefont {A.}~\bibnamefont
  {Abashian}} \emph {et~al.} (\bibinfo {collaboration} {Belle Collaboration}),\
  }\href {https://doi.org/https://doi.org/10.1016/S0168-9002(01)02013-7}
  {\bibfield  {journal} {\bibinfo  {journal} {Nucl. Instrum. Methods Phys.
  Res., Sect A}\ }\textbf {\bibinfo {volume} {479}},\ \bibinfo {pages} {117}
  (\bibinfo {year} {2002})}\BibitemShut {NoStop}%
\bibitem [{\citenamefont {Brodzicka}\ \emph {et~al.}(2012)\citenamefont
  {Brodzicka} \emph {et~al.}}]{physicsAchivementBelle}%
  \BibitemOpen
  \bibfield  {author} {\bibinfo {author} {\bibfnamefont {J.}~\bibnamefont
  {Brodzicka}} \emph {et~al.},\ }\href {https://doi.org/10.1093/ptep/pts072}
  {\bibfield  {journal} {\bibinfo  {journal} {Prog. Theor. Exp. Phys.}\
  }\textbf {\bibinfo {volume} {2012}},\ \bibinfo {pages} {04D001} (\bibinfo
  {year} {2012})}\BibitemShut {NoStop}%
\bibitem [{\citenamefont {Lange}(2001)}]{evtgen}%
  \BibitemOpen
  \bibfield  {author} {\bibinfo {author} {\bibfnamefont {D.~J.}\ \bibnamefont
  {Lange}},\ }\href
  {https://doi.org/https://doi.org/10.1016/S0168-9002(01)00089-4} {\bibfield
  {journal} {\bibinfo  {journal} {Nucl. Instrum. Methods Phys. Res., Sect A}\
  }\textbf {\bibinfo {volume} {462}},\ \bibinfo {pages} {152} (\bibinfo {year}
  {2001})}\BibitemShut {NoStop}%
\bibitem [{\citenamefont {Brun}\ \emph {et~al.}()\citenamefont {Brun} \emph
  {et~al.}}]{geant3}%
  \BibitemOpen
  \bibfield  {author} {\bibinfo {author} {\bibfnamefont {R.}~\bibnamefont
  {Brun}} \emph {et~al.},\ }\href@noop {} {\bibinfo {title} {{GEANT3}}},\
  \bibinfo {note} {{CERN Report No. DD/EE/84-1 (1987)}}\BibitemShut {NoStop}%
\bibitem [{\citenamefont {Nakano}(2002)}]{BellePID}%
  \BibitemOpen
  \bibfield  {author} {\bibinfo {author} {\bibfnamefont {E.}~\bibnamefont
  {Nakano}},\ }\href
  {https://doi.org/https://doi.org/10.1016/S0168-9002(02)01510-3} {\bibfield
  {journal} {\bibinfo  {journal} {Nucl. Instrum. Methods Phys. Res., Sect A}\
  }\textbf {\bibinfo {volume} {494}},\ \bibinfo {pages} {402} (\bibinfo {year}
  {2002})}\BibitemShut {NoStop}%
\bibitem [{\citenamefont {Feindt}\ and\ \citenamefont
  {Kerzel}(2006)}]{neurobayes}%
  \BibitemOpen
  \bibfield  {author} {\bibinfo {author} {\bibfnamefont {M.}~\bibnamefont
  {Feindt}}\ and\ \bibinfo {author} {\bibfnamefont {U.}~\bibnamefont
  {Kerzel}},\ }\href
  {https://doi.org/https://doi.org/10.1016/j.nima.2005.11.166} {\bibfield
  {journal} {\bibinfo  {journal} {Nucl. Instrum. Methods Phys. Res., Sect A}\
  }\textbf {\bibinfo {volume} {559}},\ \bibinfo {pages} {190} (\bibinfo {year}
  {2006})}\BibitemShut {NoStop}%
\bibitem [{\citenamefont {Keck}(2017)}]{fastBDTBelle}%
  \BibitemOpen
  \bibfield  {author} {\bibinfo {author} {\bibfnamefont {T.}~\bibnamefont
  {Keck}},\ }\href {https://doi.org/10.1007/s41781-017-0002-8} {\bibfield
  {journal} {\bibinfo  {journal} {Comput. Software Big Sci.}\ }\textbf
  {\bibinfo {volume} {1}},\ \bibinfo {pages} {2} (\bibinfo {year}
  {2017})}\BibitemShut {NoStop}%
\bibitem [{FW-()}]{FW-and-KSFW}%
  \BibitemOpen
  \href@noop {} {}\bibinfo {note} {{G.C.~Fox and S.~Wolfram, Phys.\ Rev.\
  Lett.\ \textbf{41}, 1581 (1978). The modified moments used in this paper are
  described in S.H.~Lee {\it et al.} (Belle Collaboration), Phys.\ Rev.\ Lett.\
  \textbf{91}, 261801 (2003).}}\BibitemShut {Stop}%
\bibitem [{\citenamefont {Asner}\ \emph {et~al.}(1996)\citenamefont {Asner}
  \emph {et~al.}}]{cleoCone}%
  \BibitemOpen
  \bibfield  {author} {\bibinfo {author} {\bibfnamefont {D.~M.}\ \bibnamefont
  {Asner}} \emph {et~al.} (\bibinfo {collaboration} {CLEO Collaboration}),\
  }\href {https://doi.org/10.1103/PhysRevD.53.1039} {\bibfield  {journal}
  {\bibinfo  {journal} {Phys. Rev. D}\ }\textbf {\bibinfo {volume} {53}},\
  \bibinfo {pages} {1039} (\bibinfo {year} {1996})}\BibitemShut {NoStop}%
\bibitem [{\citenamefont {Skwarnicki}(1986)}]{crystalball}%
  \BibitemOpen
  \bibfield  {author} {\bibinfo {author} {\bibfnamefont {T.}~\bibnamefont
  {Skwarnicki}},\ }\href@noop {} {Ph.D. thesis},\ \bibinfo  {school} {Institute
  of Nuclear Physics, Krakow} (\bibinfo {year} {1986}),\ \bibinfo {note} {the
  original Crystal Ball function is a Gaussian with a tail on the lower side.
  The function defined as a Gaussian with a tail on the upper side is used in
  this work.}\BibitemShut {Stop}%
\bibitem [{\citenamefont {Albrecht}\ \emph {et~al.}(1990)\citenamefont
  {Albrecht} \emph {et~al.}}]{ARGUSfunc}%
  \BibitemOpen
  \bibfield  {author} {\bibinfo {author} {\bibfnamefont {H.}~\bibnamefont
  {Albrecht}} \emph {et~al.} (\bibinfo {collaboration} {ARGUS Collaboration}),\
  }\href {https://doi.org/https://doi.org/10.1016/0370-2693(90)91293-K}
  {\bibfield  {journal} {\bibinfo  {journal} {Phys. Lett. B}\ }\textbf
  {\bibinfo {volume} {241}},\ \bibinfo {pages} {278} (\bibinfo {year}
  {1990})}\BibitemShut {NoStop}%
\bibitem [{\citenamefont {Hanhart}\ \emph {et~al.}(2010)\citenamefont
  {Hanhart}, \citenamefont {Kalashnikova},\ and\ \citenamefont
  {Nefediev}}]{Hanhart2010}%
  \BibitemOpen
  \bibfield  {author} {\bibinfo {author} {\bibfnamefont {C.}~\bibnamefont
  {Hanhart}}, \bibinfo {author} {\bibfnamefont {Y.~S.}\ \bibnamefont
  {Kalashnikova}},\ and\ \bibinfo {author} {\bibfnamefont {A.~V.}\ \bibnamefont
  {Nefediev}},\ }\href {https://doi.org/10.1103/PhysRevD.81.094028} {\bibfield
  {journal} {\bibinfo  {journal} {Phys. Rev. D}\ }\textbf {\bibinfo {volume}
  {81}},\ \bibinfo {pages} {094028} (\bibinfo {year} {2010})}\BibitemShut
  {NoStop}%
\bibitem [{uni()}]{unitInSecVB}%
  \BibitemOpen
  \href@noop {} {}\bibinfo {note} {{We always use $c=1$ in this
  subsection.}}\BibitemShut {Stop}%
\bibitem [{jps()}]{jpsiomegaBelle}%
  \BibitemOpen
  \href@noop {} {}\bibinfo {note} {{K. Abe {\it et al.} (Belle Collaboration),
  \href{https://arxiv.org/abs/hep-ex/0505037}{arXiv:hep-ex/0505037}}}\BibitemShut
  {NoStop}%
\bibitem [{\citenamefont {del Amo~Sanchez}\ \emph {et~al.}(2010)\citenamefont
  {del Amo~Sanchez} \emph {et~al.}}]{jpsiomegaBABAR}%
  \BibitemOpen
  \bibfield  {author} {\bibinfo {author} {\bibfnamefont {P.}~\bibnamefont {del
  Amo~Sanchez}} \emph {et~al.} (\bibinfo {collaboration} {\BABAR
  Collaboration}),\ }\href {https://doi.org/10.1103/PhysRevD.82.011101}
  {\bibfield  {journal} {\bibinfo  {journal} {Phys. Rev. D}\ }\textbf {\bibinfo
  {volume} {82}},\ \bibinfo {pages} {011101} (\bibinfo {year}
  {2010})}\BibitemShut {NoStop}%
\bibitem [{\citenamefont {Ablikim}\ \emph {et~al.}(2019)\citenamefont {Ablikim}
  \emph {et~al.}}]{jpsiomegaBESIII}%
  \BibitemOpen
  \bibfield  {author} {\bibinfo {author} {\bibfnamefont {M.}~\bibnamefont
  {Ablikim}} \emph {et~al.} (\bibinfo {collaboration} {BESIII Collaboration}),\
  }\href {https://doi.org/10.1103/PhysRevLett.122.232002} {\bibfield  {journal}
  {\bibinfo  {journal} {Phys. Rev. Lett.}\ }\textbf {\bibinfo {volume} {122}},\
  \bibinfo {pages} {232002} (\bibinfo {year} {2019})}\BibitemShut {NoStop}%
\bibitem [{BRj()}]{BRjpsipipi}%
  \BibitemOpen
  \href@noop {} {}\bibinfo {note} {Our own average using the most recent
  measurements from Belle, \BABAR and LHCb~\cite{BelleBRJpsipipi2011,
  BABARBRJpsipipi2008, LHCbBRJpsipipi2020}}\BibitemShut {NoStop}%
\bibitem [{RXJ()}]{RXJpsipipi}%
  \BibitemOpen
  \href@noop {} {}\bibinfo {note} {{Our own average of ${\cal B}(B^0\to
  X(3872)K^0)/{\cal B}(B^+ \to X(3872)K^+)$ using the most recent measurements
  from Belle and \BABAR~\cite{BelleBRJpsipipi2011, BABARBRJpsipipi2008},
  $0.48\pm0.13$}}\BibitemShut {NoStop}%
\bibitem [{\citenamefont {Choi}\ \emph {et~al.}(2011)\citenamefont {Choi} \emph
  {et~al.}}]{BelleBRJpsipipi2011}%
  \BibitemOpen
  \bibfield  {author} {\bibinfo {author} {\bibfnamefont {S.-K.}\ \bibnamefont
  {Choi}} \emph {et~al.} (\bibinfo {collaboration} {Belle Collaboration}),\
  }\href {https://doi.org/10.1103/PhysRevD.84.052004} {\bibfield  {journal}
  {\bibinfo  {journal} {Phys. Rev. D}\ }\textbf {\bibinfo {volume} {84}},\
  \bibinfo {pages} {052004} (\bibinfo {year} {2011})}\BibitemShut {NoStop}%
\bibitem [{\citenamefont {Aubert}\ \emph
  {et~al.}(2008{\natexlab{b}})\citenamefont {Aubert} \emph
  {et~al.}}]{BABARBRJpsipipi2008}%
  \BibitemOpen
  \bibfield  {author} {\bibinfo {author} {\bibfnamefont {B.}~\bibnamefont
  {Aubert}} \emph {et~al.} (\bibinfo {collaboration} {\BABAR Collaboration}),\
  }\href {https://doi.org/10.1103/PhysRevD.77.111101} {\bibfield  {journal}
  {\bibinfo  {journal} {Phys. Rev. D}\ }\textbf {\bibinfo {volume} {77}},\
  \bibinfo {pages} {111101} (\bibinfo {year} {2008}{\natexlab{b}})}\BibitemShut
  {NoStop}%
\end{thebibliography}%

\end{document}